\documentclass[useAMS,usenatbib]{mn2e}
\usepackage{graphicx}
\usepackage{txfonts}
\usepackage{enumerate}
\usepackage{xcolor} 

\newcommand{\gamete}{\textsc{GAMETE/QSOdust}}
\newcommand{\msun}{M$_\odot$}

\newcommand{\mh}{H$_2$ } 

\title[seed BHs]{From the first stars to the first black holes}

\author[Valiante et al.]{Rosa Valiante$^{1}$\thanks{E-mail:
rosa.valiante@oa-roma.inaf.it}, Raffaella Schneider$^{1}$, Marta Volonteri$^{2}$,
Kazuyuki Omukai$^{3}$\\
$^{1}$INAF - Osservatorio Astronomico di Roma, via di Frascati 33, 00040, Monteporzio Catone, 
Italy\\
$^{2}$ CNRS, UMR 7095, Institut d’Astrophysique de Paris, F-75014, Paris, France \\
$^{3}$ Astronomical Institute, Tohoku University, Aoba, Sendai 980-8578, Japan
}

\begin{document}

\date{Accepted . Received }

\pagerange{\pageref{firstpage}--\pageref{lastpage}} 
\pubyear{2016}
\maketitle

\begin{abstract}
The growth of the first super massive black holes (SMBHs) at $z > 6$ is
still a major challenge for theoretical models. If it starts from
black hole (BH) remnants of Population III stars (light seeds 
with mass $\sim 100$ M$_\odot$) it requires super-Eddington
accretion. An alternative route is to start from heavy seeds
formed by the direct collapse of gas onto a $\sim 10^5$ M$_\odot$
BH. Here we investigate the relative role of light and heavy
seeds as BH progenitors of the first SMBHs. We use the 
cosmological, data constrained semi-analytic model \gamete
\, to simulate several independent merger histories of $z > 6$
quasars. Using physically motivated prescriptions to form light and heavy 
seeds in the progenitor galaxies, 
we find that the formation of a few heavy seeds (between 3 and 30 in our reference
model)  enables the Eddington-limited  growth of 
SMBHs at $z > 6$. This conclusion depends sensitively on the
interplay between chemical, radiative and mechanical feedback effects,
which easily erase the conditions that allow the suppression
of gas cooling in the low metallicity gas 
($Z < Z_{\rm cr}$ and $J_{\rm LW} > J_{\rm cr}$). 
We find that heavy seeds can not form if dust cooling triggers gas fragmentation above a
critical dust-to-gas mass ratio (${\cal D} \ge {\cal D}_{\rm cr}$). 
In addition, the relative importance of light and heavy seeds
depends on the adopted mass range for light seeds, as this dramatically
affects the history of cold gas along the merger tree, by both SN and
AGN-driven winds. 

\end{abstract}

\section{Introduction}

Super Massive Black Holes (SMBHs), powering the most luminous quasars 
($>10^{47}$ erg s$^{-1}$) at redshift $z>6$, are among the most 
intriguing and puzzling astronomical objects observed in the early Universe.
Observational campaigns are pushing the high redshift frontier closer and 
beyond the reionization epoch, expanding the census of high redshift quasars. 

The most distant quasar observed so far is ULAS J1120+0641, at $z\sim 7$, 
in which the central engine is a black hole (BH) with a mass of 
$M_{\rm BH}=2.0^{+1.5}_{-0.7}\times 10^9$ \msun \, \citep{Mortlock11}, 
already in place when the Universe was as old as $\sim 700$ Myr. 
Recently, \citet{Wu15} discovered an ultraluminous quasar at $z\sim 6.3$ 
($\sim 4\times 10^{14}$ L$_\odot$) hosting a massive BH of 
$(1.2\pm 0.19)\times 10^{10}$ \msun, presumably accreting close to the 
Eddington rate.

The existence of $10^9-10^{10}$ \msun \, BHs in the early Universe 
(\citeauthor{Fan01} \citeyear{Fan01, Fan04}; 
\citeauthor{deRosa11} \citeyear{deRosa11, deRosa14} and 
references therein) poses a challenge to theoretical models aimed to 
explain the formation and growth of such massive objects. 
Many efforts have been done so far in order to
unveil the nature of their progenitor seed BHs,
how and when these seeds form and how they can grow so rapidly, in less than 
$\sim 1$ Gyr, up to few billion solar masses and more.

Different scenarios 
for BH formation have been proposed so far (see e.g. \citealt{Rees78, 
Volonteri10}, \citealt*{VolonteriBellovary12}
for comprehensive reviews).
BH seeds with $M_{\rm BH} \sim 10^2$ M$_\odot$ are predicted to form as end products 
of massive, metal-poor Population III (Pop~III) stars (e.g. \citealt{Abel02, Heger03};
\citealt*{MadauRees01}; \citealt{Yoshida08, Latif13b, Hirano14}).
Runaway collisions of massive stars during the gravitational collapse of 
the core of compact star clusters can lead to the formation  of  intermediate mass 
BHs with $M_{\rm BH} \sim 10^3-10^4$ M$_\odot$ (e.g. \citealt{Omukai08}; \citealt*{Devecchi09};
\citealt{Katz15}). In addition, the fast merging of stellar mass BHs in a cluster
has been proposed as a possible way to give rise to more massive seeds
(\citealt{Davies11, Lupi14}). Finally, the formation of more massive, $10^4-10^6$ \msun, seed BHs 
is predicted to occur via direct collapse of dense, 
metal poor gas clouds in halos with virial temperatures $T_{\rm vir}\geq 10^4$ K which 
are exposed to a strong \mh photodissociating flux (e.g. \citealt{BL03, Begelman06, 
SpaansSilk06}; \citealt*{IO12}; \citealt{Inayoshi14, Ferrara14}).
Low metallicity and the suppression of \mh molecules formation are fundamental requirements 
for avoiding gas cooling, cloud fragmentation and thus star formation. 

It has been suggested that the less massive seeds (Pop III remnants and collapsed 
stellar clusters) would require a continuous gas accretion close to or above the 
Eddington limit in order to grow up to few billion solar masses in less than
$\sim 1$ Gyr. 
For example, \citet{Johnson13} show that BH seeds as massive as
$10^5$ \msun, are required if sub-Eddington accretion and a large radiative efficiency
($\epsilon_{\rm r}\geq 0.1$) are assumed (see also \citealt*{VSD15}). 

Although the conditions in which they can form are met only in very rare 
environments (e.g. \citeauthor{Hosokawa12} \citeyear{Hosokawa12, Hosokawa13}; 
\citealt{Inayoshi14}; \citealt*{IH14}, \citealt{Sugimura15, Yue14})
direct collapse BHs (DCBHs) have been proposed as a viable 
scenario to explain  high redshift SMBHs.
The formation mechanism, characteristic mass and relevant processes in the direct 
collapse scenario have been widely investigated in the literature 
(e.g. \citeauthor{Latif13a} \citeyear{Latif13a, Latif13b, Latif14b}; 
\citealt{Johnson12, Sugimura14}; 
\citeauthor{Agarwal14} \citeyear{Agarwal12, Agarwal14, Agarwal15};
\citealt{LatifVolonteri15}, \citeauthor{Glover15a} \citeyear{Glover15a, Glover15b}).

The work presented in this paper is similar in spirit to what has been done 
by \citet{Petri12} who presented a semi-analytic model for the assembly of 
high-$z$ SMBHs along a merger history, starting from DCBHs of $10^5$ \msun \, 
and stellar mass BHs of $10^2$ \msun.
They show that the final BH mass assembled via both gas accretion and BH-BH mergers, 
strongly depends on the fraction of halos hosting DCBHs in the merger tree. 
A $\sim 10^{10}$ \msun \, BH can be obtained if this fraction reaches $100\%$.
However, these authors do not follow the chemical evolution of the host galaxies, 
and in particular of the metallicity and dust-to-gas ratio of the interstellar medium.

The aim of this paper is to investigate the relative role of the less massive, Pop III remnant,
seed BHs (light seeds) and of the most massive DCBH seeds (heavy seeds) in the formation 
and evolution of the first SMBHs, taking into account the BH-host galaxy co-evolution.
To this aim we adopt an improved version of the semi-analytic code \gamete \,
which has been successfully used to investigate the evolutionary scenarios of
high redshift quasars at $z>5$ (\citealt{V11,V12,V14}). In particular, together with
the mass of the BH, the model well reproduce the properties of the quasars host galaxies 
such as the star formation rate, the mass of gas, metals and dust.

As in \citet{V11} we select as our target the quasar SDSS J1148+5251, 
observed at redshift $z=6.4$. This is one of the best studied object at
high redshift, hosting a BH mass of $(2-6)\times 10^9$ \msun \, 
(\citealt{Barth03, Willott03}). The other main observed properties of this 
quasar are summarized in \citet{V11} and \citet{V14}. 

The paper is organized as follows. 
In section 2 we present our approach introducing the model \gamete \, 
and presenting in details the new features implemented for this work.
The results are presented in section 3 where we study 
the redshift evolution of the BH mass and the birth environment of the seeds
predicted by the reference model. In section 4 we discuss the dependence of 
the results on the model assumptions and parameters, and we summarize the 
conclusions in section 5.

\section{Summary of the model}
Here we briefly introduce \gamete, the semi-analytic model adopted for 
this study, focusing on the new features implemented to investigate the 
nature of the first seed black holes.
We refer the reader to \citeauthor{V11} (\citeyear{V11,V14}) for a full description 
of the code.

\gamete\, is a data constrained model aimed at the study of the formation and evolution 
of the first quasars and their host galaxies.  
It is based on the semi-analytic merger tree model \textsc{GAMETE}
which was originally developed by \citet{SS07} to investigate the early evolution
of the Milky Way and later applied to the Milky Way dwarf satellites \citep{SS08,SS09,SS15},
to investigate their contribution to the reionization and metal enrichment  history of the 
Local Group \citep{SS14}, and to explore the connection of Damped Lyman-$\alpha$ systems
with local dwarfs \citep{SS12}.
A two-phase interstellar medium (ISM) version of \textsc{GAMETE} was successfully 
adopted as a stellar archaeology tool to investigate the origin of metal-poor 
low-mass stars in the Milky Way (\citealt{deBen14}, \citeauthor{deBen14} in prep).

In its present version, \gamete \, enables us to investigate the co-evolution
of nuclear black holes and their host galaxies, following their star formation
histories and the enrichment of their ISM with metals and dust. 
The gas reservoir inside each galaxy is 
regulated by processes of star formation, BH growth and feedback.

We assume that the first SMBHs, observed at $z>5$, reside in dark 
matter (DM) halos of $M_{\rm h} = 10^{13}$ \msun \, 
Thus, we first reconstruct the merger tree of such massive halos, decomposing them 
into their progenitors, backward in time, following a binary Monte Carlo approach 
based on the Extended Press-Schechter theory (see \citealt{V11} and references 
therein for details): at each timestep, a halo of mass $M_{\rm h}$ can either lose mass or 
lose mass and fragment into two less massive ($\leq M_{\rm h}/2$) progenitors.

Along the merger history of the DM progenitors, we follow the gradual evolution of 
the central SMBH and its host galaxy, via both mass accretion and mergers.
We define as major mergers halos merging with mass ratio $\mu>1/4$, where $\mu$ is 
the ratio of the less massive halo over the most massive companion.
In each galaxy, the star formation rate (SFR) is assumed to be proportional to the 
available gas mass and the efficiency at which the gas is transformed into stars is 
enhanced during major mergers (see \citeauthor{V11} \citeyear{V11,V14}). 

Following \citet{V11}, we assume
that in major mergers pre-existing BHs merge in symbiosis with their 
host galaxies and form of a new, more massive BH. 
In minor mergers BHs are unable to spiral in on short timescales as the 
merger time scale is on the order of the Hubble time or longer (see e.g. 
\citealt{TH09}). As a result, the least massive BH of the merging pair remains as a 
satellite and we do not follow its evolution.  

We assume Eddington-limited BH growth. The
accretion rate is described by the Bondi-Hoyle-Lyttleton (BHL) formula
where we introduce a free parameter, $\alpha_{\rm BH}$, which is commonly adopted
to quantify the increased density in the inner regions around the BH \citep{diMat05}.
A value of $\alpha_{\rm BH} = 50 $ is fixed to match the observed SMBH mass of J1148.
A fraction of the energy released by supernova explosions and BH accretion
is converted into kinetic energy of the gas in the host galaxy, thus driving gas outflows 
in the form of winds. In our previous study \citep{V12} we show that the BH-host galaxy co-evolution is 
regulated by quasar feedback, with SN-driven winds providing a negligible 
contribution to the mass outflow rate, in good agreement with observations of 
outflowing gas in J1148 \citep{M12,Cicone15}.
The AGN-driven wind efficiency, $\epsilon_{\rm w,AGN} =  2.5 \times 10^{-3}$ is fixed to match the observed 
gas mass in the host galaxy of J1148 (see Section~2.7).

Finally,  the ISM of each progenitor galaxy is progressively enriched with metals 
and dust produced by Asymptotic Giant Branch (AGB) stars and Supernovae (SNe) 
according to their stellar evolutionary time scales. 
Here we adopt the improved version of the chemical evolution module of \gamete \,
presented in \citet{V14}, where we consistently follow the evolution of metals and dust
 taking into account the dust life-cycle in a two-phase ISM. 
In hot diffuse gas, dust grains can be destroyed by SN shocks while in cold, dense 
clouds - where stars form - dust grains can grow by accretion of gas-phase heavy elements.

In the following sections we describe the new features introduced in the model
for the purpose of the present study.

\subsection{Resolving mini-halos}
\label{sec:DMhalos}

\begin{figure*}
\includegraphics [width=8.0cm]{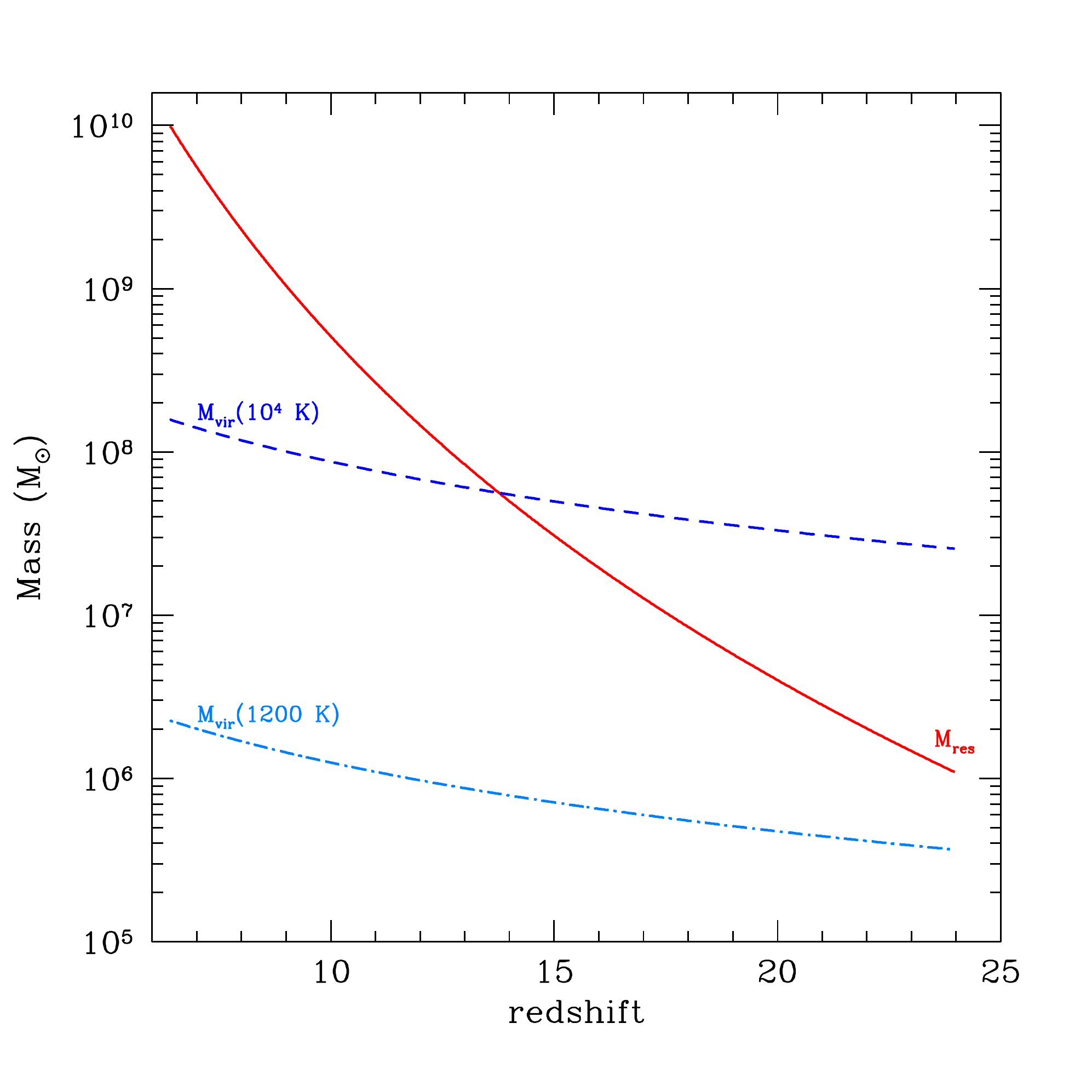}
\includegraphics [width=8.0cm]{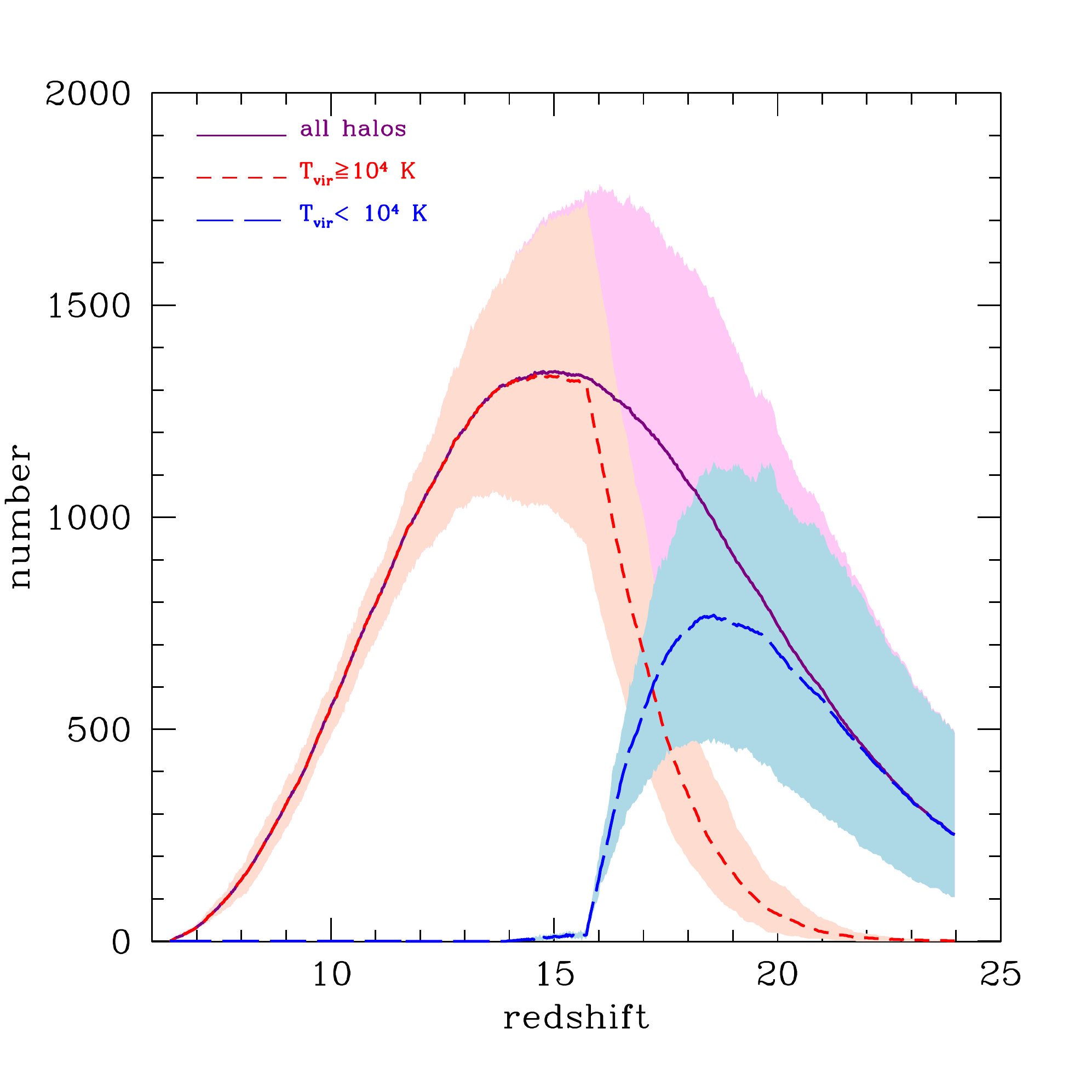}
\caption{Left panel: redshift evolution of the resolution mass adopted in the merger
tree simulations (solid red line) and of the dark matter halo masses corresponding to
a virial temperature of $1200$ K (azure dot-dashed line) and $10^4$ K (blue dashed line).
Right panel: the number of dark matter halos as a function of redshift averaged over
10 independent realizations of the merger tree (solid line). Long and short-dashed lines
show the separate contributions of mini-halos and Lyman-$\alpha$ halos (see text).
Shaded regions represent the minimum and maximum values at each redshift.} 
\label{fig:fig1} 
\end{figure*}

The first collapsed objects where the gas is able to cool and form stars have small masses, 
$M_{\rm h} \sim 10^{5-6}$ \msun, and virial temperatures $T_{\rm vir}< 10^4$~K  
(see e.g. \citealt{Bromm13} for a recent review).  To resolve these so-called
{\it mini-halos}, we simulate the merger trees adopting a minimum mass of:
\begin{equation}
M_{\rm res}(z_i) = 10^{-3} \, M_{\rm h}(z_0)\,\Big(\frac{1+z_i}{1+z_0}\Big)^{-7.5}
\label{eq:mres}
\end{equation}
\noindent
where $M_{\rm h}(z_0)=10^{13}$ M$_\odot$ is the host DM halo at redshift $z_0=6.4$.
The redshift evolution of this resolution mass is shown in the left panel of 
Fig.~\ref{fig:fig1}. This choice enables us to simulate different 
realizations of the merger tree resolving high-$z$ mini-halos in a relatively short 
computational time. 

In mini-halos, which we define here as dark matter halos with $1200~{\rm K} \le T_{\rm vir}< 10^4$ K, 
the primordial gas can cool via rotational transitions of \mh (e.g. \citealt{HTL96}).
Dark matter halos whose virial temperature exceeds the threshold for efficient atomic 
line cooling,  $T_{\rm vir} \geq 10^4$~K, are instead referred to as {\it Lyman$-\alpha$} (Ly$\alpha$) halos. 
The number of mini-halos and Ly$\alpha$ halos averaged over 10 different merger
tree realizations is shown as a function of redshift in the right panel of 
Fig.~\ref{fig:fig1}. 
As expected, at $z \gtrsim 17$ mini-halos represent the dominant population among dark matter
progenitors. Their number decreases at lower redshift down to $z \sim 14$, below which 
the halo population is completely dominated by more massive systems.
This is a consequence of the redshift evolution of the 
assumed resolution mass which exceeds the minimum mass of Ly$\alpha$ halos at these redshifts\footnote{In other words, we are note able to resolve mini-halos at redshift 
$z<14$ due to the chosen resolution mass threshold. 
This does not affect our results because at these $z$ radiative feedback has already 
suppressed star and BH formation in all halos below a virial 
temperature of $10^4$ K. This will be discussed in section \ref{sec:results}.}.

\subsection{UV radiation}
\label{sec:UV}

The radiation emitted by stars and accreting BHs gradually builds up a cosmic ultraviolet (UV) 
background.
Since we reconstruct the merger history of a single biased, high density region at 
$z>6$, here we refer to UV background as the cumulative emission
coming from all the progenitor galaxies of the $M_{\rm h} = 10^{13}$ \msun \, DM halo 
within its comoving volume at the turn-around radius, $V_{\rm com} = 50 \, \rm Mpc^3$.

Radiative feedback effects have a fundamental role in the history of star and BH seed formation. 
Photons in the Lyman$-$Werner (LW) band, $[11.2 - 13.6]$ eV, can easily dissociate 
\mh molecules, suppressing cooling and star formation in metal-poor 
mini-halos (e.g. \citealt{HRL97a, HRL97b}, \citealt{ON99}, \citealt{Omukai01}, 
\citealt{Mach01}). Even a moderate LW flux can lead to an increase of the minimum mass required for DM 
halos to host star formation (see Appendix A). 

The increased gas temperature in photo-ionized regions leads to an increase of the 
cosmological Jeans mass. As a result, gas accretion onto low-mass dark matter halos
is suppressed in ionized regions, while the internal gas in existing low-mass halos will be
photo-evaporated (e.g. \citealt{BL99}, \citealt{Shap04}, \citealt{SM13}).

The cumulative flux (in units of erg cm$^{-2}$ s$^{-1}$ Hz$^{-1}$ sr$^{-1}$) at the 
observed frequency $\nu_{\rm obs}$ and redshift $z_{\rm obs}$ is computed as (e.g. 
\citealt{HM96}):

\begin{equation}
J(\nu_{\rm obs},z_{\rm obs}) = \frac{(1+z_{\rm obs})^3}{4\pi} 
\int_{z_{\rm obs}}^{z_{\rm max}} dz \, c \, \Big| \frac{dt}{dz}\Big|
\, \epsilon(\nu_{z},z) \,  e^{-\tau_{\rm H_2}(\nu_{\rm obs},z_{\rm obs},z)} 
\label{eq:Jlw}
\end{equation} 
where $\tau_{\rm H_2}$ is the \mh optical depth in the LW band and $\epsilon(\nu_{z},z)$ 
is the comoving emissivity, namely the monochromatic luminosity per unit comoving
volume (erg s$^{-1}$ Hz$^{-1}$ cm$^{-3}$), in the LW band at redshift $z$. 
The redshift $z_{\rm max}$ represents the highest redshift from which a LW photon emitted
by a source at $z>z_{\rm obs}$ can reach the observer at $z_{\rm obs}$ before being redshifted
at lower frequencies, outside the LW range, into a H Lyman resonance line. 
In the dark screen approximation, this 
redshift can be defined as $(1+z_{\rm max})/(1+z_{\rm obs}) = \nu_{i}/\nu_{\rm obs}$ (see e.g. 
\citealt{HRL97}, \citealt{HAR00}) being $\nu_{i}$ the first Lyman line frequency 
above the observed one.

In general, $\tau_{\rm H_2}$ depends on \mh number density, on the line profile and 
on the probability that the molecule is dissociated after a transition. It can reach values 
$\tau_{\rm H_2}\geq 3$ (\citealt{CFA00}, \citealt{RGS01})
leading to a reduction in the LW background flux of about one order of magnitude.
Following \citet{Ahn09} we compute the intergalactic absorption averaged over the LW band
using the modulation factor described by the fitting formula: 
\begin{equation}
 e^{-\tau_{\rm H_2}} = \left\{
 \begin{array}{lr}
    1.7 \, e^{-(r_{\rm cMpc}/116.29\alpha)^{0.68}}-0.7 \,\,\,\,\,  {\rm if} \, \,\, r_{\rm cMpc}/\alpha \le 97.39\\
    0  \,\,\,\,\,  {\rm if}\,  \,\, r_{\rm cMpc}/\alpha > 97.39
 \end{array}
 \right.
\label{eq:absorp}
\end{equation}
\noindent
where $r_{\rm cMpc}$ is the distance between the emitting source, at redshift $z$,
and the observer at redshift $z_{\rm obs}$, expressed in units of comoving Mpc:
\begin{equation}
 r_{\rm cMpc} = -\int^{z}_{z_{\rm obs}} \frac{c dz}{H(z)} 
 \label{eq:separation}
\end{equation}
\noindent
with 
$H(z) = H_{0}[\Omega_{\rm M}(1+z)^3 + \Omega_\Lambda]^{1/2}$. The scaling factor $\alpha$ in 
eq.~(\ref{eq:absorp}) is defined as:
\begin{equation}
 \alpha = \left ( \frac{h}{0.7} \right)^{-1} \, \left( \frac{\Omega_{\rm M}}{0.27} \right )^{-1/2} 
          \, \left ( \frac{1+z}{21} \right )^{-1/2}.
 \label{alpha}
\end{equation}
The resulting average attenuation of the UV flux increases with increasing comoving ratio 
$r_{\rm c Mpc}/\alpha$, approaching zero when $r_{\rm cMpc}=97.39\alpha$ which is the maximum 
distance from which the observer can see LW photons emitted by a source at redshift $z$, 
the so-called LW horizon (see Fig.~3 of \citealt{Ahn09}).

In what follows, we call $J_{\rm LW}$ the LW background flux in units 
of $10^{-21}$ erg cm$^{-2}$ s$^{-1}$ Hz$^{-1}$ sr$^{-1}$  and we compute it at the central
frequency of the LW band using eqs.~(\ref{eq:Jlw}) --  (\ref{alpha}).

Following \citet{SS14}, the time evolution of the filling factor of ionized regions is computed as:
\begin{equation}
\dot{Q}_{\rm HII}(z)= f_{\rm esc} \dot{n}_{\gamma}/n_{\rm H} - \alpha_{B} \, C \, n_{\rm H} \, (1+z)^3 Q_{\rm HII}
 \label{eq:QHII}
\end{equation}
\noindent
where $f_{\rm esc} =  0.1$ is the escape fraction of ionizing photons,
$\dot{n}_{\gamma}=\sum_{i} \dot{N}_{\gamma,i}/V_{com}$ is the
total production rate of ionizing photons per unit volume 
summed over all the emitting sources, 
$n_{\rm H} = X_{\rm H} \, n_{\rm IGM}$ is the comoving hydrogen number density in the integalactic medium (IGM), $n_{\rm IGM}$ is the IGM gas number density and $X_{\rm H}=0.76$ is the 
hydrogen mass fraction. 
In the right-hand side of eq.~(\ref{eq:QHII}), $\alpha_{B}=2.6 \times 10^{-13}$ cm$^3$ s$^{-1}$ is 
the hydrogen recombination rate and $C = 3$ is the clumping factor. 

At each given redshift, the total LW emissivity and ionizing photon rate are computed 
summing over all the emitting sources, both stars and 
accreting BHs.
For Pop~III stars, we use the mass-dependent emissivities
given by \citet{Schaerer02} for $Z=0$ stars with no mass loss (see Table 4 and Table 6 of the original paper). 
For Pop II/I stars we compute the metallicity and age-dependent  emissivities using \citet{BC03} 
population synthesis model. We assume that the stars form in a single burst with a Salpeter
Initial Mass Function (IMF) in the mass range $[0.1-100]$ \msun.
For accreting BHs, we compute the LW and ionizing photons production rates by modeling the Spectral Energy 
Distribution (SED) as a classic multicolor disk spectrum up to $k T_{\rm max} \sim 1 \, {\rm keV} \, (M_{\rm BH}/ \rm M_\odot)^{-1/4}$
\citep{SS1973}, and a non-thermal power-law component with spectral slope $L_\nu \propto \nu^{-\alpha}$, with $\alpha \approx 2$
at higher energies \citep{SS1973,SOS04}.

\subsection{Star formation rate} 
\label{sec:SF}

In each progenitor galaxy, the star formation rate is computed as:
\begin{equation}\label{eq:sfr}
{\rm SFR } = f_{\rm cool} \, M_{\rm ISM} \, (\epsilon_{\rm quies}+\epsilon_{\rm burst})/\tau_{\rm dyn} 
\end{equation}
\noindent
where $M_{\rm ISM}$ is the gas mass, $\epsilon_{\rm quies}+\epsilon_{\rm burst}$ is the
total star formation efficiency accounting for both quiescent and merger-driven episodes of star formation. 
Following \citet{V11}, we take $\epsilon_{\rm quies}=0.1$ and $\epsilon_{\rm burst} = 8$ for equal mass
mergers, with a modulation that depends on the mass ratio of the merging pairs
(see eq.~11 and Table~2 in \citealt{V11}). Finally, the quantity $f_{\rm cool}$ quantifies the 
reduced cooling efficiency of mini-halos with respect to Ly$\alpha$ halos. Hence, we assume $f_{\rm cool} = 1$
in progenitor systems with $T_{\rm vir} \ge 10^4$~K whereas in mini-halos this parameter quantifies the
mass fraction of gas that can cool in one dynamical time and it depends on the virial temperature,
redshift, gas metallicity and intensity of the LW background. The computation of $f_{\rm cool}$ 
is described in Appendix A.\\

\noindent 
\textit{Photo-heating feedback.}\\
To account for the effects of the increased gas temperature in photo-ionized regions,
we assume that star formation is suppressed, i.e. 
$(\epsilon_{\rm quies}+\epsilon_{\rm burst})=0$, in halos with virial temperature below the 
temperature of the IGM. Hence, we neglect the hydrodynamic
response of the gas (see Sobacchi \& Mesinger 2013) and we assume a
feedback model where star formation is suppressed instantaneously when $T_{\rm vir}<T_{\rm IGM}$.
The mean IGM temperature is computed taking into account the volume filling factor of
ionized regions, 
$T_{\rm IGM} = Q_{\rm HII} \, T_{\rm reio} + (1-Q_{\rm HII}) \, T_{\rm gas}$ where $T_{\rm reio} = 2 \times 10^4$ K is 
the assumed post-reionization temperature and $T_{\rm gas} = 170 \, (1+z)^2$. \\

\noindent 
\textit{Photo-dissociating feedback}.\\
Suppression of $\rm H_2$ cooling and star formation in mini-halos due to
photo-dissociation by LW photons is taken into account through the 
parameter $f_{\rm cool}$ in eq.~(\ref{eq:sfr}), whose calculation
is presented  in Appendix A. Depending on the halo virial temperature,
redshift, gas metallicity and intensity of the LW background, we compare 
the cooling time and the free-fall time and quantify the mass fraction of
gas that is able to cool and form stars. We find that in the presence
of a LW background, the cooling efficiency is rapidly suppressed in
mini-halos. In fact, when $J_{\rm LW} \lesssim 1$, 
$f_{\rm cool} \neq 0$ only in mini-halos at $z \gtrsim 20$ or if the gas is already
metal-enriched to $Z \gtrsim 0.1$ Z$_\odot$.  For more details, we refer the
reader to Appendix A and Figs.~\ref{fig:A1}-\ref{fig:A4}.

\subsection{Stellar initial mass function} 
\label{sec:popIII}
Currently, there are no direct observational constraints on the initial mass function of the 
first generation of stars. Theoretical studies do not yet provide a firm determination of
the stellar mass spectrum emerging from the first star forming regions (see e.g. 
\citealt{Bromm13} and \citealt{Glover13} for comprehensive reviews).
Recent numerical studies suggest that - depending on their formation environment -
Pop III stars can have masses varying from 10s to 1000s \msun, with a distribution 
that peaks around few tens to few hundreds solar masses 
(e.g. \citealt{Hosokawa11,Hirano14,Susa14,Hirano15}).

\begin{figure}
\includegraphics [width=8.0cm]{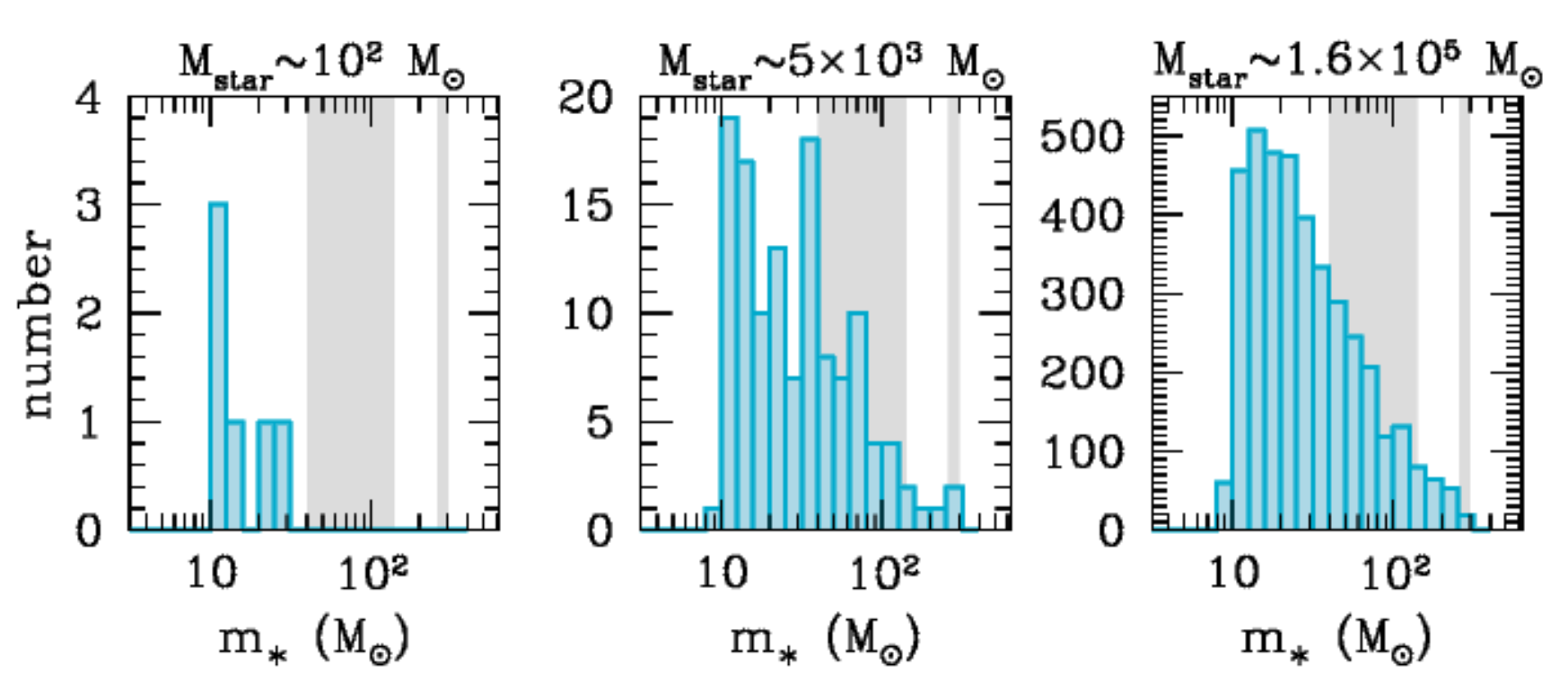}
\caption{Examples of the mass distribution of Pop~III stars emerging from stochastic
sampling of the IMF in three different halos (see text). The total stellar mass formed is 
$10^2 \, \rm M_\odot$ (left panel),
$\sim 5 \times 10^3 \, \rm M_\odot$ (middle panel) and $\sim 1.6\times 10^5 \, \rm M_\odot$ 
(right panel).
Grey shaded regions indicate the mass range leading to BH remnants.} 
\label{fig:randomIMF} 
\end{figure}

Following \citet{deBen14}, we assume that Pop III stars form according to a Larson IMF \citep{Larson1998},
\begin{equation}
\Phi(m_\ast) \propto m_\ast^{\alpha-1} \, \, e^{-m_\ast/m_{\rm ch}},
\end{equation}
\noindent
with $\alpha = - 1.35$, $m_{\rm ch} = 20$ M$_\odot$ and $10$ M$_\odot \le m_\ast \le 300$ M$_\odot$. 
During each star formation episode, we stochastically sample the IMF until we reach the total
stellar mass formed\footnote{We have tested that the IMF
is fully reconstructed in the $[10-300]\, \rm M_\odot$ mass range when the total stellar mass formed is 
$M_{\rm star}>10^6 \, \rm M_\odot$.}(see also \citealt{deBennassuti15}). In Fig.~\ref{fig:randomIMF} we show three examples 
of the mass distribution of Pop~III stars emerging from a star formation episode where the
total stellar mass formed is $M_{\rm star} \sim 10^2 \, \rm M_\odot$  (left panel), 
$\sim 5\times 10^3 \, \rm M_\odot$ (middle panel) and $\sim 1.6\times 10^5 \, \rm M_\odot$ (right panel). 
The stellar population shown in the left panel is representative of the conditions that apply in small
mass mini-halos. Only 6 stars are formed with masses in the range $[10  - 30]\, \rm M_\odot$. In larger
mass halos, as shown in the middle and right panels, the mass range that is populated is extended
toward larger stellar masses.

When the metallicity in star forming regions is $Z_{\rm ISM} \ge Z_{\rm crit}$, where
$Z_{\rm crit}$ is the critical metallicity for low-mass star formation
 (\citeauthor{Schneider02} \citeyear{Bromm01, Schneider02, Schneider03}), we assume that 
Pop~II stars form in the stellar mass range $[0.1-100] \, \rm M_\odot$ according to a Larson 
IMF with $m_{\rm ch} = 0.35 \, \rm M_\odot$. In what follows, we adopt $Z_{\rm crit} = 10^{-3.8}$ Z$_\odot$
and we discuss the impact of assuming a dust-driven transition at lower $Z_{\rm crit}$ \citep{Omukai05,Schneider12a}.
 
\subsection{Light black hole seed formation} 
\label{sec:lightSeeds}

Light BH seeds form as remnants of Pop~III stars. Here we assume that stars with 
masses in the range $[40 - 140]\, \rm M_\odot$ and $\ge 260\, \rm M_\odot$ do not 
explode as SNe and directly collapse to BHs \citep{Heger02}. The number 
and masses of BH remnants depend on the frequency with which these 
mass ranges are sampled when Pop~III stars form and stochastically populate
the IMF (see the previous section).
In the two halos shown in the middle and right
panels of Fig.~\ref{fig:randomIMF}, the stochastic sampling selects 124 and 3900 stars.
The shaded regions in the same figure indicate the mass range leading to BH remnants.
In small mini-halos, light BH seeds are very rare (see left panel).

The subsequent evolution of newly formed BHs depends on their mass. 
As discussed by \citet{Volonteri10}, lighter BHs are not expected to settle at the 
center but rather wander through the host galaxy, interacting with stars. For this
reason,  we select as a light BH seed the most 
massive BH remnant\footnote{ In the three examples shown in Fig.~\ref{fig:randomIMF}, 
we do not assign any light BH seed to the population represented by the left panel
and only one light BH seed in the other cases, taken to be the most massive BH remnant
among the 2 (15) BHs of $\sim [260 - 300] \, \rm M_\odot$ formed
in the middle (right) panel.}.
We discuss the implications of this assumption in section \ref{sec:discussion}. 

\subsection{Heavy black hole seed formation}
\label{sec:heavySeeds}

Fragmentation of gas clouds, and thus star formation, is prevented in Ly$\alpha$ halos
($T_{\rm vir}\geq 10^4$ K) in which the ISM metallicity is sub-critial ($Z_{\rm ISM}<Z_{\rm cr}$) 
and the LW background is strong enough to photo-dissociate $\rm H_2$ \citep{Omukai08}. 
The latter condition is usually expressed as $J_{\rm LW}>J_{\rm cr}$,
where $J_{\rm cr}$ is the critical value in units of $10^{-21}$ erg cm$^{-2}$ s$^{-1}$Hz$^{-1}$ sr$^{-1}$.
If all the above conditions are simultaneously satisfied, the collapse proceeds almost 
isothermally thanks to atomic H line cooling, avoiding fragmentation into smaller clumps. 
This process leads to the formation of a single BH with mass in the range 
$[10^{4}-10^{6}] \, \rm M_\odot$,
that we call heavy BH seed, in some cases through an intermediate  phase of super-massive star formation 
(see e.g. \citealt{Hosokawa12,Hosokawa13} for more details). 
Recently, \citet{Ferrara14} investigated the mass spectrum of heavy BH seeds and found that their masses 
range between  $\sim 5\times 10^4\, \rm M_\odot$ \, to $\sim 2\times 10^6 \, \rm M_\odot$ 
(see also \citealt{VB10}). 

Following these studies, we assume that heavy BH seeds form with an average mass of 
$10^5 \, \rm M_\odot$ in Ly$\alpha$ halos with sub-critical metallicity 
and super-critical LW background. 

The exact value of $J_{\rm cr}$ is still a matter of debate. 
Its value depends on whether sufficient $\rm H_2$ to cool the gas within
a free-fall time is formed before it
is collisionally dissociated at $\sim 10^4 \rm \,cm^{-3}$, and depends on the spectral energy distribution of the sources of radiation 
\citep{Omukai01,OhHaiman02, BL03, Omukai08, Agarwal12, Latif14b, Sugimura14, Sugimura15, Agarwal15}.
In addition, Ly$\alpha$ halos can be exposed to intense local radiation, which exceeds the background 
level, in biased, dense regions of the Universe, and close to star forming galaxies \citep{Dijkstra08,TH09,Dijkstra14}. 
Additional complications come when \mh self-shielding is taken into account  \citep{Shang10, Hartwig15} and 
when the presence of X-ray or ionizing radiation increase the free electron fraction, favoring the formation
of $\rm H_2$ \citep{IO11,Yue14,Johnson14,IT15}.  As a result, values of $J_{\rm cr}$ between $\sim 30$ and $\sim 10^4$
have been proposed and used to estimate the number density of heavy seeds.
In particular, $J_{\rm cr} > 500-10^3$ and up to $10^{4} - 10^{5}$ is always required to enable the direct 
collapse mechanism in 3D numerical simulations to produce supermassive stars with mass 
$10^4-10^5 \, \rm M_\odot$ \citep{Latif14a, Regan14, LatifVolonteri15}.
Here we adopt a reference value of $J_{\rm cr}=300$ and we discuss the implications of this 
assumption in section \ref{sec:discussion}. 

\begin{figure}
\includegraphics [width=8.0cm]{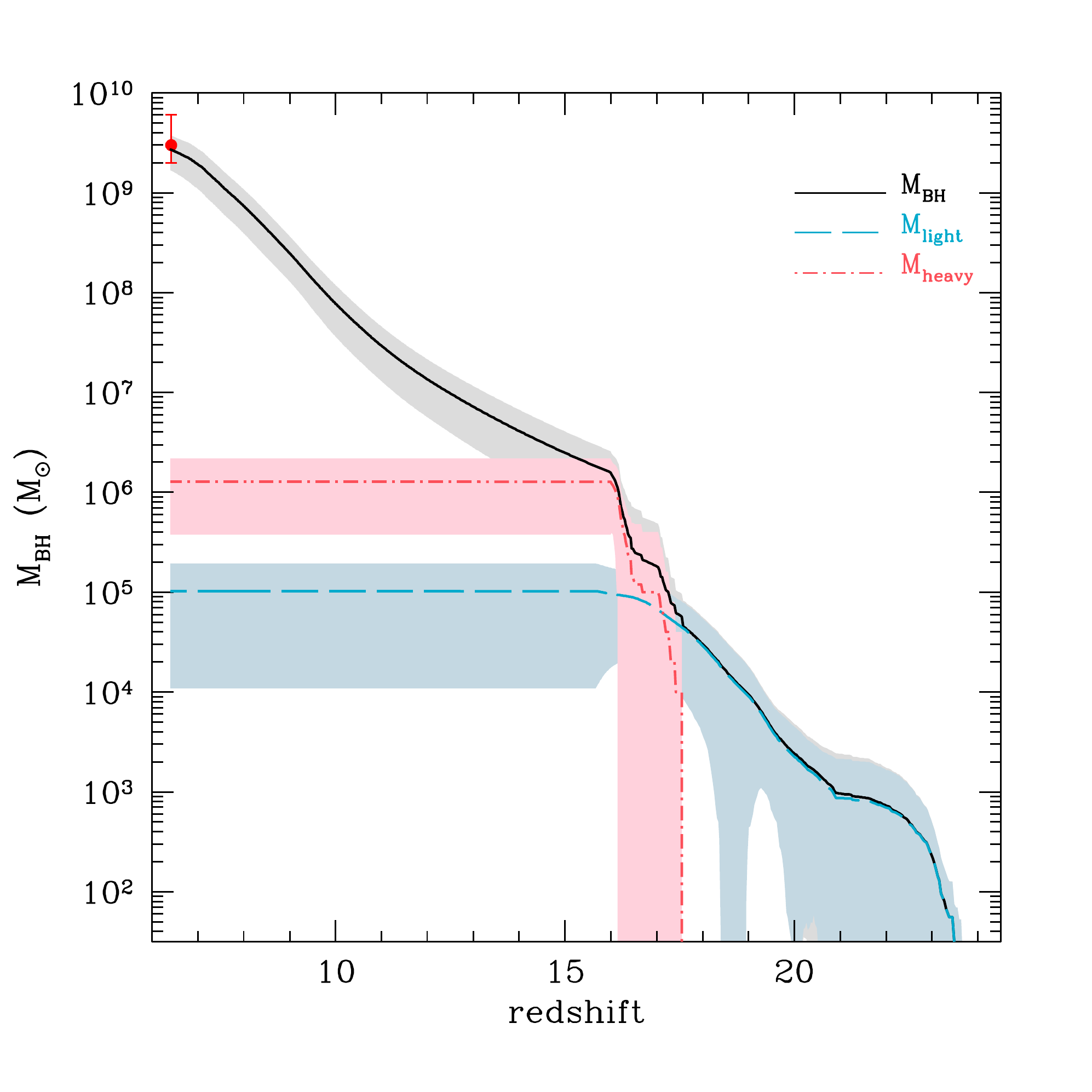}
\caption{Black hole mass as a function of redshift. 
  The black solid line represents the total
  mass in BHs growing by gas accretion and mergers with other BHs 
  while the blue dashed and the red dot-dashed lines show the contribution
  of the light and heavy seed BHs, respectively (without gas accretion).
  Each line is the average over 10 different merger tree realizations with
  the shades indicating $1\sigma$ dispersion.} 
\label{fig:bhevo} 
\end{figure} 
\begin{figure*}
\hspace{-1cm}
\includegraphics [width=6cm]{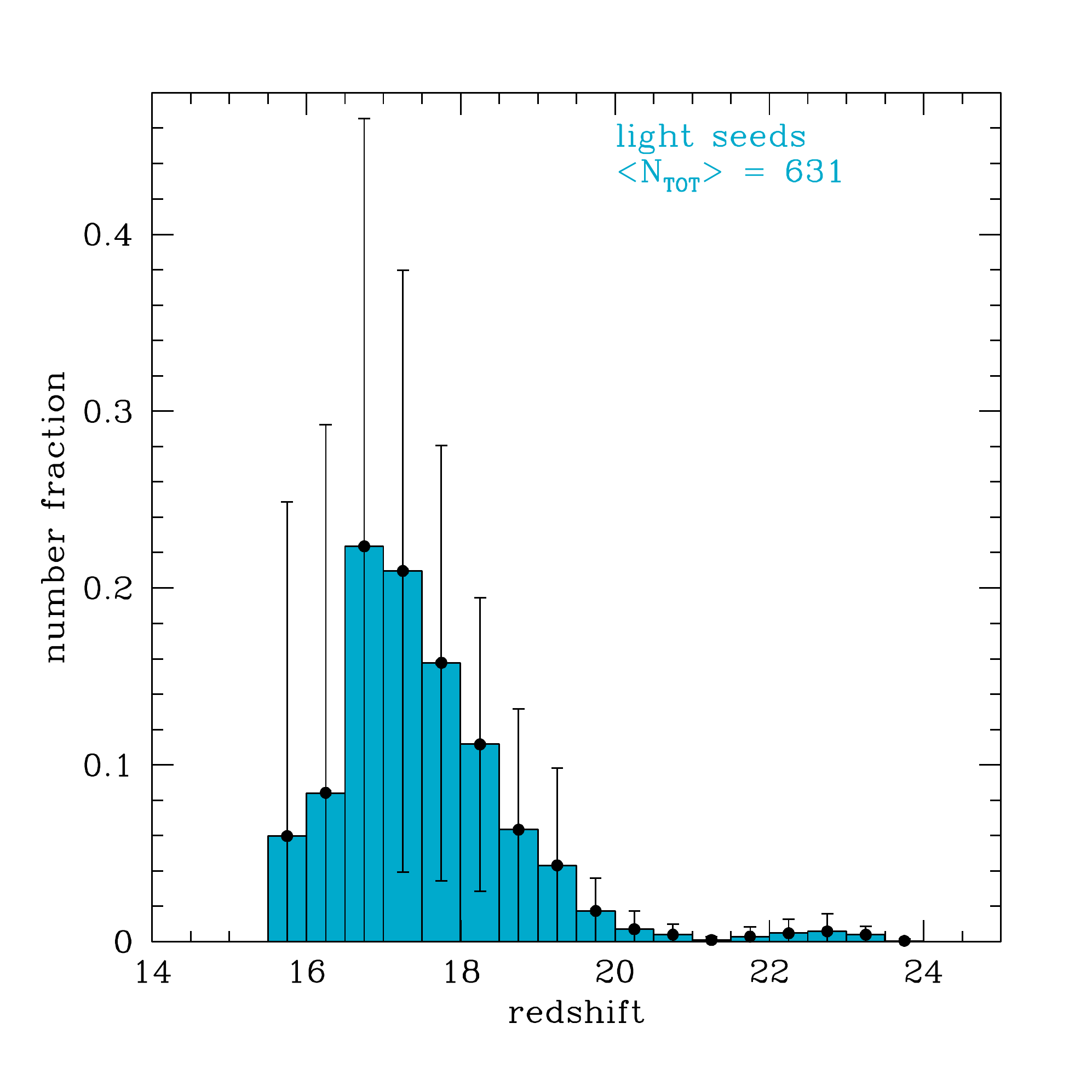}
\includegraphics [width=6cm]{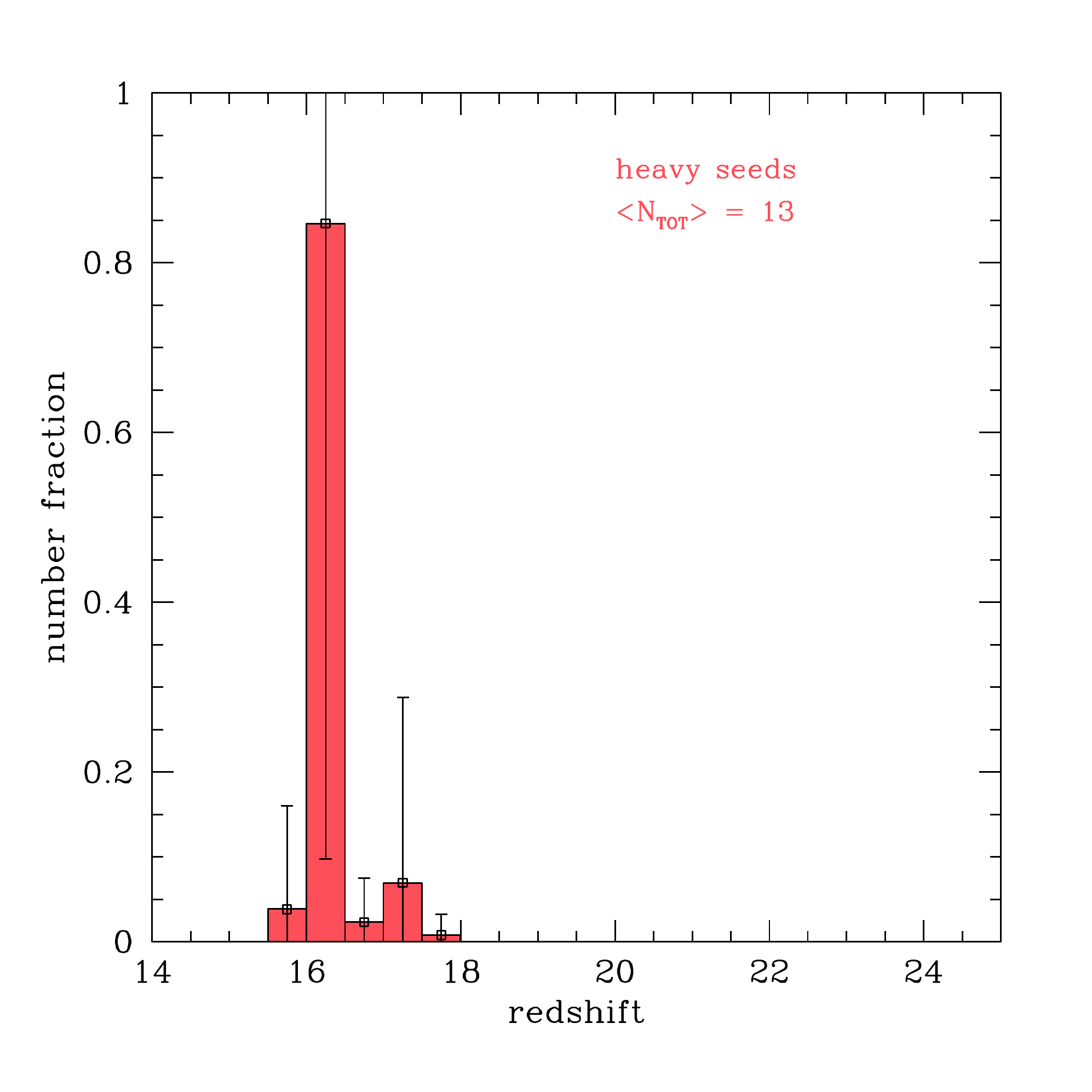}
\includegraphics [width=6cm]{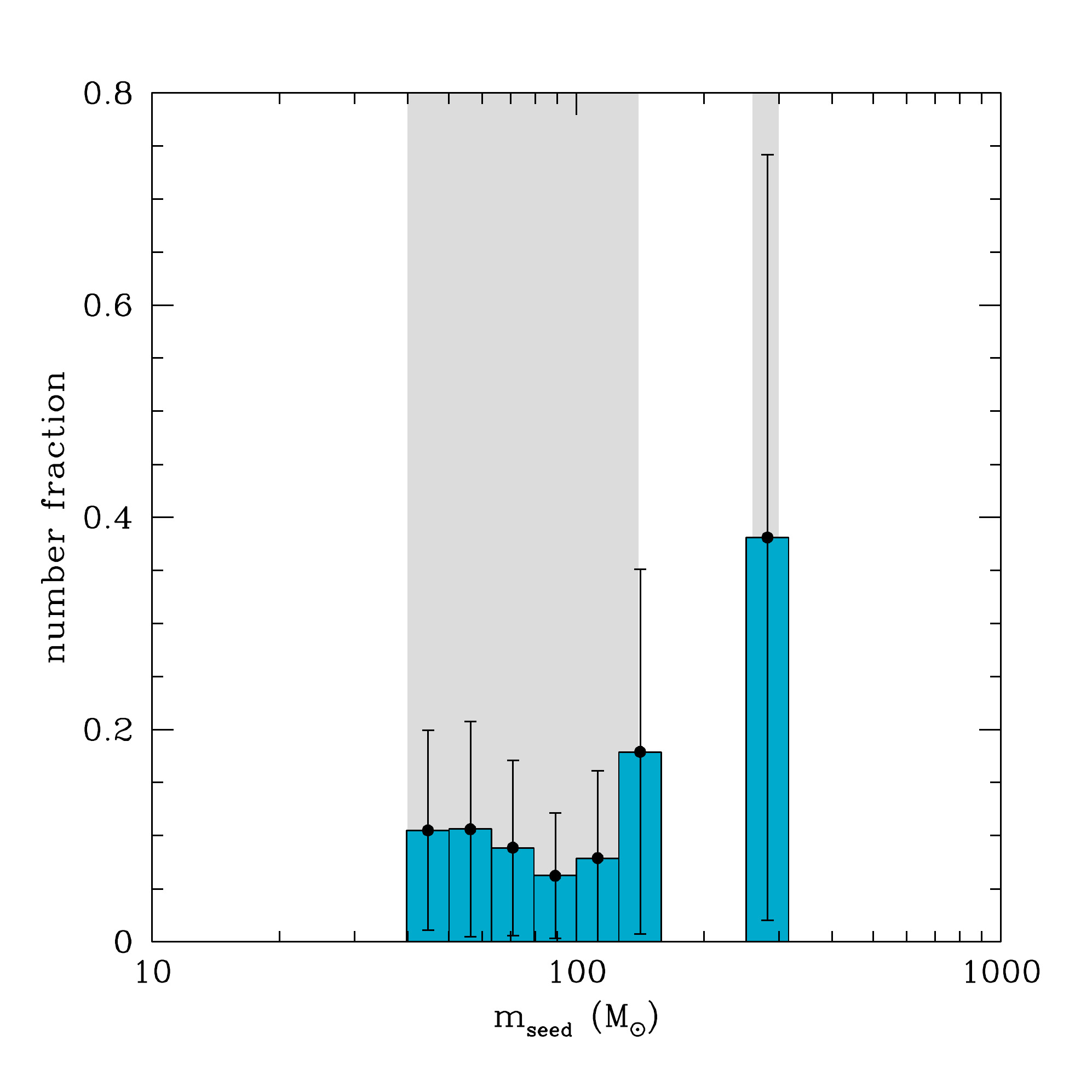}
\caption{Distribution of formation redshift of light (left panel) and heavy seeds (central panel). The
right panel shows  the distribution of birth masses for light seeds only as heavy seeds are all assumed
to have a mass of $10^5 \, \rm M_\odot$. Histograms and data points show the average over 10 different 
merger tree realizations with $1-\sigma$ error bars.} 
\label{fig:histogLS} 
\end{figure*} 

\section{Results} 
\label{sec:results}

In this section we present the main results of our study. In the same
spirit of \citet{V11,V14}, we follow the formation
of a SMBH with a mass of  $\sim (2-6)\times 10^9$ \msun \, at redshift $z=6.4$, 
similar to the one expected to power the bright quasar J1148 \citep{Barth03, Willott03, deRosa11}. 
 In what follows, we present the results
averaged over 10 independent realizations of the merger tree of a 
$10^{13}$ \msun \, DM halo. However, in order to explore the dependence
of some results on the merger history, we also discuss the properties of individual merger trees.


\subsection{Evolution of the black hole mass}
\label{sec:BHseeds}

In Fig.~\ref{fig:bhevo} we show the predicted evolution of the total BH mass in a 
merger tree as a function of redshift.
For each merger tree, we consider the contribution of BH progenitors to the total BH mass at each redshift.
We classify as BH progenitors only those BHs which do not become satellites at any stage of the 
merger tree and whose mass will be inherited by the final SMBH at $z = 6.4$\footnote{For each
merger tree, we follow backward in time the evolution of the SMBH. At each minor merger event, we cut the branch of the tree
of the lighter, satellite progenitor BH and we only follow the branch of the most massive one. At each major merger event, we continue to
follow both branches of the progenitor BHs. This procedure allows us to reconstruct a-posteriori the sample of  BH progenitors whose masses 
directly contribute to the final SMBH mass at $z = 6.4$.}.
The free parameters of the model have been selected to reproduce a SMBH mass of 
$\sim 3\times 10^9 \, \rm M_\odot$ at $z = 6.4$, in good agreement with that expected for quasar J1148 (red 
data point in Fig.~\ref{fig:bhevo}). 
In the same figure we also show the separate contribution of light (blue dashed line) and 
heavy (red dot-dashed line) BH seeds to the total BH mass at different epochs. 
\newline

\begin{figure*}
\hspace{-1 cm}
\includegraphics [width=6.0cm]{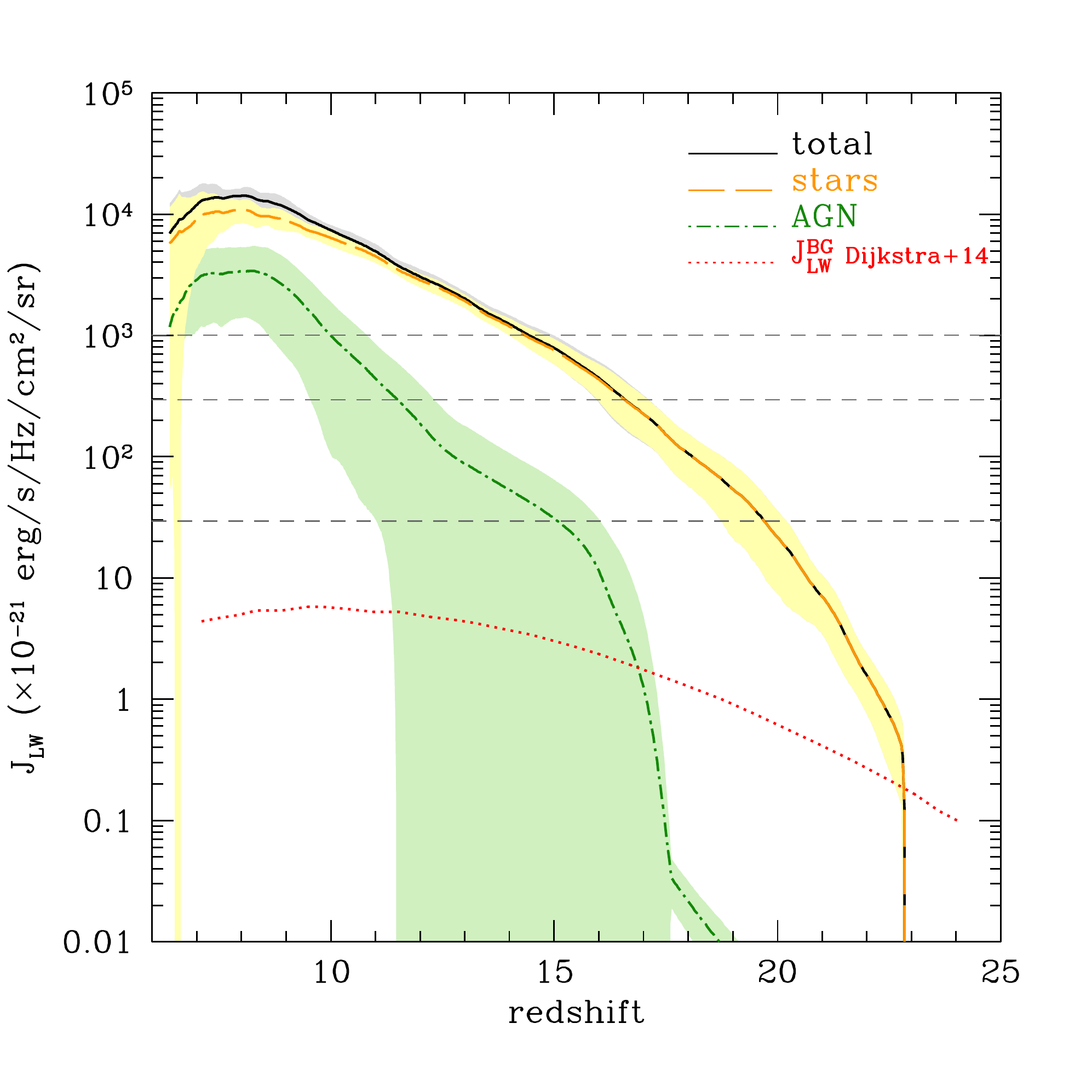}
\includegraphics [width=6.0cm]{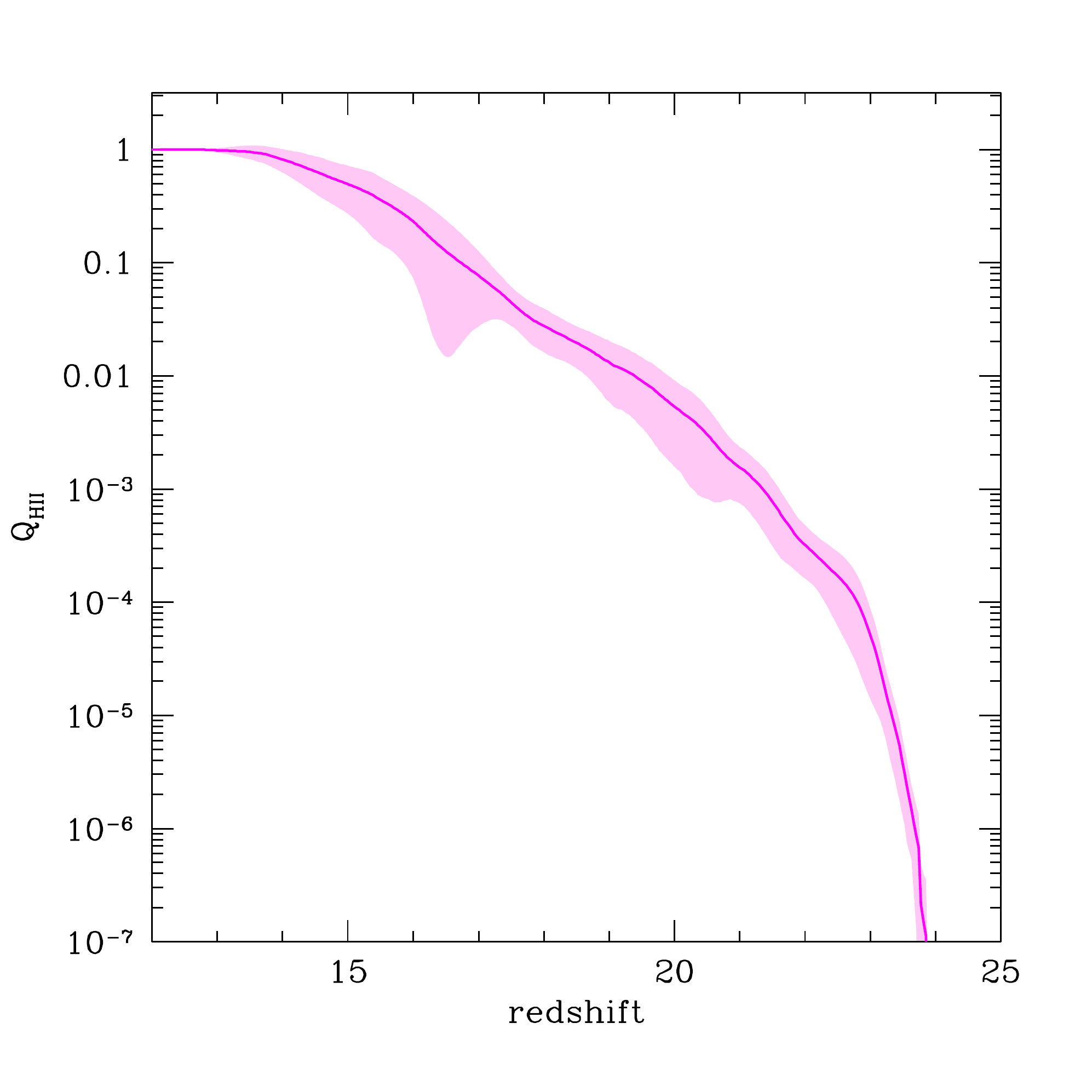}
\includegraphics [width=6.0cm]{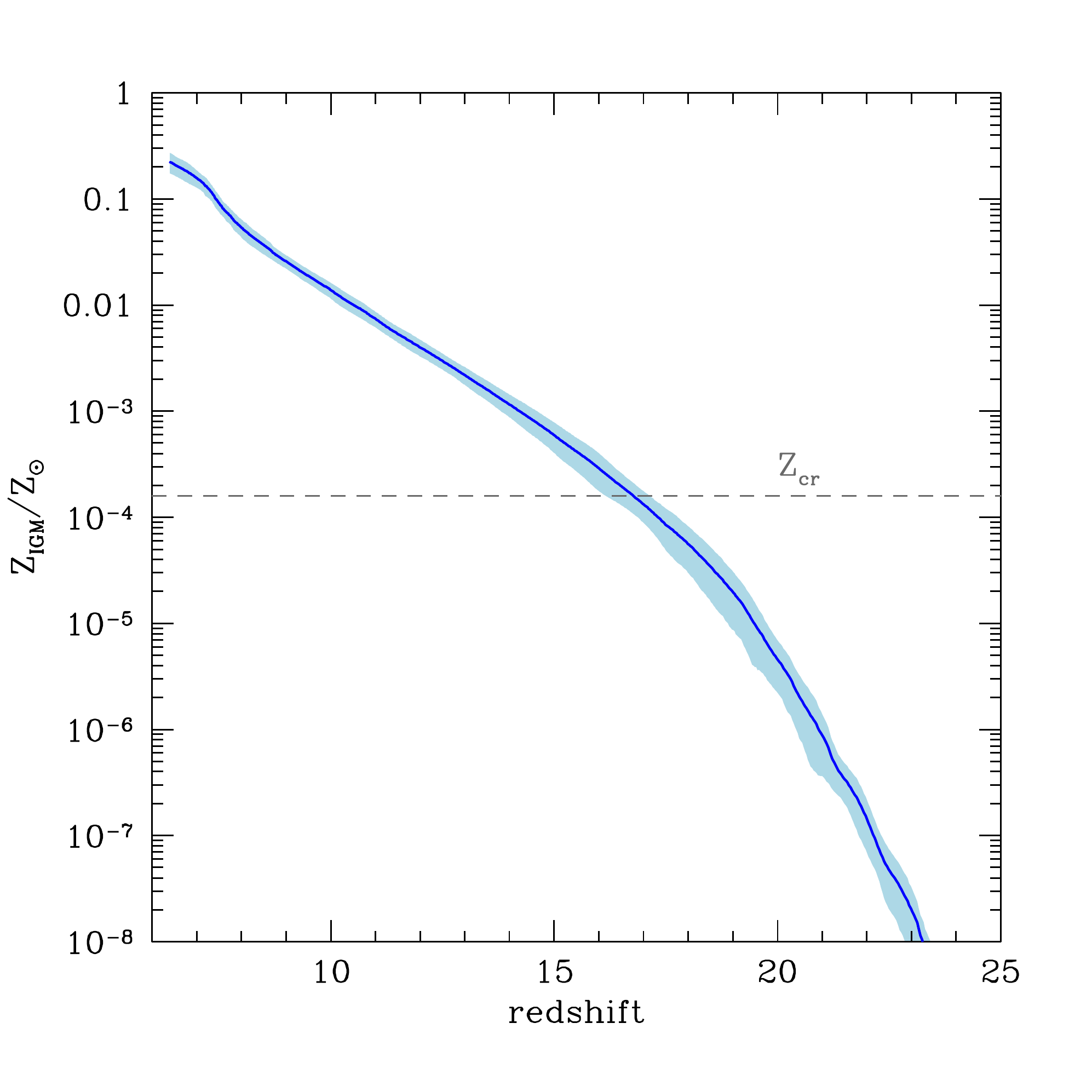}
\caption{
\textit{Left panel:} redshift evolution of the total Lyman Werner background flux
(black solid line). Green dot-dashed and yellow dashed lines indicate the contribution of 
accreting BHs (AGNs) and stars, respectively. 
The three horizontal lines indicate three different 
critical values $J_{\rm cr}=30, 300, $ and $10^3$ (see text for details). For comparison,
we also show the cosmic mean LW background predicted by Dijkstra et al. (2014, dotted
line).
\textit{Middle panel:} Volume filling factor of ionized hydrogen regions (HII) as a 
function of redshift. 
\textit{Right panel:} redshift evolution of the metallicity of the IGM in which halos are 
embedded. The horizontal line shows the critical metallicity 
for Pop III/II transition adopted in the reference model. 
In all panels, lines indicate the average values over 10 different merger tree 
realizations with shades representing the $1-\sigma$ dispersion. 
}
\label{fig:feedback} 
\end{figure*} 

\noindent \textit{Light seeds-dominated regime}.
At high redshift ($z \gtrsim 18$), BH growth is dominated by 
the formation of light seeds. Their rate of formation
is strongly regulated by photo-dissociating feedback which inhibits
Pop~III star formation in mini-halos.
The total mass from BH light seeds rapidly grows in time, 
reaching, on average, a maximum value of $\sim 10^5 \, \rm M_\odot$
at $z \sim 15.5$, below which their formation is suppressed by
metal enrichment. 
\newline

\noindent \textit{Heavy seeds-dominated regime}.
Heavy seeds start to form at redshift $z \lesssim 18$ and dominate
the evolution of the BH mass for a brief but significant period of time.
In fact, they rapidly grow in number and by $z \sim 15.5$
their contribution to the total BH mass is, on average, $\sim 1.3\times 10^6$ \msun,
more than one order of magnitude larger than that of light BH seeds. Not
surprisingly, the rise of heavy seeds marks the fall of
light seeds. In fact, Ly$\alpha$ halos with $Z < Z_{\rm cr}$ either form Pop~III stars,
hence light BH seeds (when $J_{\rm LW} < J_{\rm cr}$), or form heavy seeds 
(when $J_{\rm LW} \ge J_{\rm cr}$). 
\newline

\noindent \textit{Accretion-dominated regime}.
At $z \lesssim 16$, BH growth is dominated by gas accretion. In fact, at this epoch
the progenitor galaxies are all enriched to $Z \ge Z_{\rm cr}$ preventing the formation
of both light and heavy seeds. Overall, gas accretion provides the dominant contribution
to the final SMBH mass at $z = 6.4$, in agreement with \citet{V11}\footnote{
In \citet{V11} gas accretion dominates the evolution of the BH mass at  $z \lesssim 11$.
In the present model gas accretion starts to dominate at an earlier 
redshift. The difference with \citet{V11} is due to the different merger histories (which now include 
mini-halos) and BH seeding prescription.}. This is a consequence of the strong BH mass dependence
of the BHL accretion law, which leads to run-away BH growth.
\newline

In our reference model, a total of $\sim 4800$ light and 
$\sim 100$ heavy seeds are formed, on average, at $z \gtrsim 16$. 
However, only $\sim 13\%$ of these seeds ($\sim 620$ light and $\sim 13$ heavy) 
are BH progenitors, because a dominant fraction is lost along minor branches of the merger tree 
and become satellites. 

In Fig.~\ref{fig:histogLS}  we show the BH progenitor
formation redshifts and birth masses.  
Light BH seeds start forming at $z \sim 24$ although their number increases considerably 
at $z \lesssim 20$, with a peak at $z \sim 17$ followed by a rapid decline. This redshift
distribution reflects the properties of their birth environments. At
the highest $z$, light BH seeds form in mini-halos, whose star formation efficiency is 
low and prone to photo-dissociating feedback. 
Their birth mass distribution shows that the largest 
number of light BH seeds is concentrated in the most massive bin, with $\sim 240$ BHs ($\sim 40\%$
of the total) with mass in the range $[260 - 300]\, \rm M_\odot$, while the remaining are almost 
equally distributed between $40$ and $140 \, \rm M_\odot$ (see the right panel).
On the other hand, heavy seed BH progenitors form, on average, over a very narrow
redshift range, at $15.5 \lesssim z \lesssim 18$, with a peak at $z \sim 16.5$ that is 
slightly shifted with respect to that of light seeds, followed by a sharp decline.  This
sudden appearance and decline of heavy seeds is a consequence of their tight
birth environmental conditions which are satisfied only by a relatively small number
of halos and over a very limited period of time, as it will be clarified in the following
section.

\subsection{Birth environment of SMBH seeds}
\label{sec:environment}

The results presented in the previous section show how the mass growth of 
SMBHs at $z > 6$ depends on a complex interplay between 
radiative and chemical feedback processes that shape the birth environment of light and
heavy BH seeds.  Since SMBHs at $z > 6$ form in biased regions of the Universe, 
the intensity of the LW background, the volume 
filling factor of ionized regions and the gas metallicity, which set the relative strength
of feedback processes, are expected to be different from the cosmic mean values
at the same redshift. 

The left panel of Fig.~\ref{fig:feedback} shows the relative contribution of
accreting BHs (AGNs, green dot-dashed line) and stellar emission (yellow dashed line)
to the LW background (black solid line). At all redshifts, the LW
emission is dominated by star formation. The intensity of the LW background
increases very rapidly, exceeding values of $J_{\rm cr} = 30, 300$, and  
$10^3$ (marked by the horizontal lines), on average,  at $z \sim 20, 16.5,$ and $15$,
with some dispersion among different merger histories (see section \ref{sec:singelMTs}).
In the same figure we also show, for comparison, the cosmic mean LW background 
predicted by \citet{Dijkstra14}.

The central panel of Figure \ref{fig:feedback} shows the evolution of the volume filling factor
of ionized regions. We find that  $Q_{\rm HII} \sim 1$ at $z \lesssim 14$, consistent with the
expectations from the rapid increase of the UV background intensity.

Finally, in the right panel of Figure \ref{fig:feedback} we present the redshift evolution 
of the metallicity of the IGM, the medium in which all halos are embedded. 
This metallicity, $Z_{\rm IGM}$, increases as mechanical feedback, in the form of 
galaxy-scale winds driven by the SNe and AGNs, ejects metal-enriched gas 
out of the galaxies, enriching the surrounding medium. 
The horizontal line indicates the critical metallicity for low-mass star formation
that we have adopted in the reference model, $Z_{\rm cr} = 10^{-3.8}$ Z$_\odot$. 
The average IGM metallicity exceeds this critical value at $z \lesssim 17$, with
some dispersion among different merger histories.

\begin{figure}
\includegraphics [width=8.0cm]{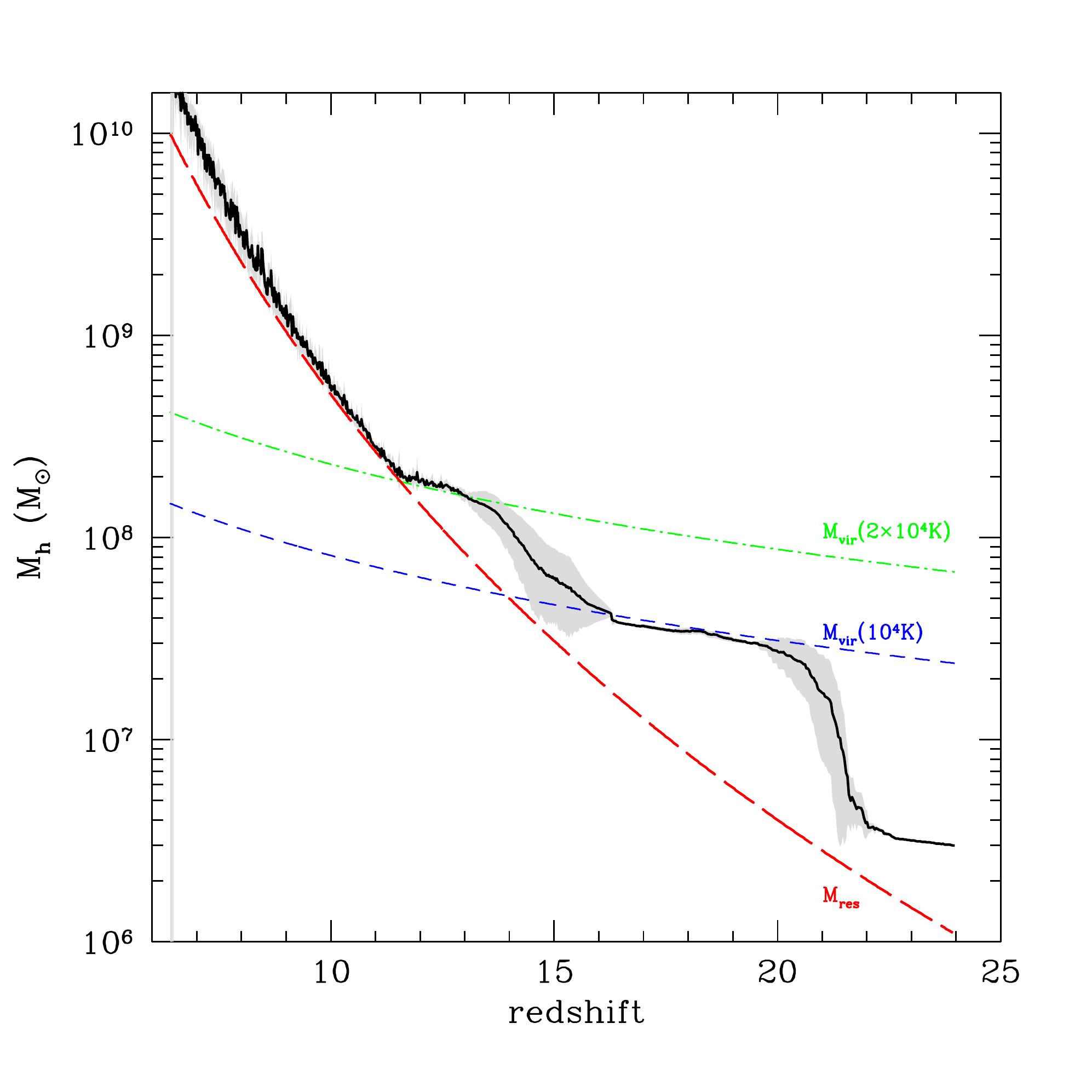}
\caption{Average minimum halo mass for star forming as a function of redshift 
(black solid line). 
The line indicates the average over 10 different merger histories and the
shaded region shows the $1-\sigma$ dispersion. 
For comparison, we also show the resolution mass of the merger trees (red long-dashed line), the 
minimum mass of  Ly$\alpha$ halos (with $T_{\rm vir} = 10^4$~K, blue short-dashed line) and 
the adopted minimum mass for star formation in ionized regions (with $T_{\rm vir} = 2\times 10^4$~K, green dot-dashed line).}
\label{fig:massLimit} 
\end{figure}

The effects of radiative feedback on the environment
where light and heavy BH seeds form is summarized in Fig.~\ref{fig:massLimit}, which shows the redshift
evolution of the minimum halo mass for star formation (black solid line). 
At high redshifts the minimum
halo mass rapidly increases as a consequence of photo-dissociating feedback, reaching
the minimum mass of Ly$\alpha$ halos (short-dashed line) already at $z \sim 20$, on average. Hence, as
it was anticipated in section~\ref{sec:BHseeds}, the dominant fraction of light BH seeds
form in Ly$\alpha$ halos 
which are less vulnerable to photo-dissociating feedback. Between
$16 \lesssim z \lesssim 20$ Ly$\alpha$ halos with masses $M_{\rm h} \sim (3 - 5)\times 10^7 \, \rm M_\odot$ and
sub-critical metallicity can either form
light or heavy BH seeds depending on the intensity of $J_{\rm LW}$. When $z \lesssim 16$
the minimum mass for star formation increases as a consequence of photo-heating
feedback and achieve the adopted minimum mass for star formation in ionized regions, 
$M_{\rm h} \gtrsim 2 \times 10^8 \, \rm M_\odot$ (dot-dashed line), by $z \sim 13$, when the IGM
is fully ionized. The figure also shows that the minimum mass for star formation depends on the
adopted resolution mass only at $z \lesssim 11$, when the epoch of BH seed formation is already
terminated. Hence, the results are independent of the mass resolution of the merger trees. 
  
Finally, to quantify the effect of chemical feedback on heavy BH seeds formation, we
compute their occurence ratio, defined as the number of progenitor halos which satisfy
the conditions $J_{\rm LW} > J_{\rm cr}$ and $Z < Z_{\rm cr}$ divided by the
number of progenitor halos with $J_{\rm LW} > J_{\rm cr}$. When averaged 
over 10 independent merger histories, we find the occurrence 
ratio at $z > 15$ to be $\sim 5\%$, meaning that chemical 
feedback plays a dominant role.
 
\begin{figure}
\includegraphics [width=8.0cm]{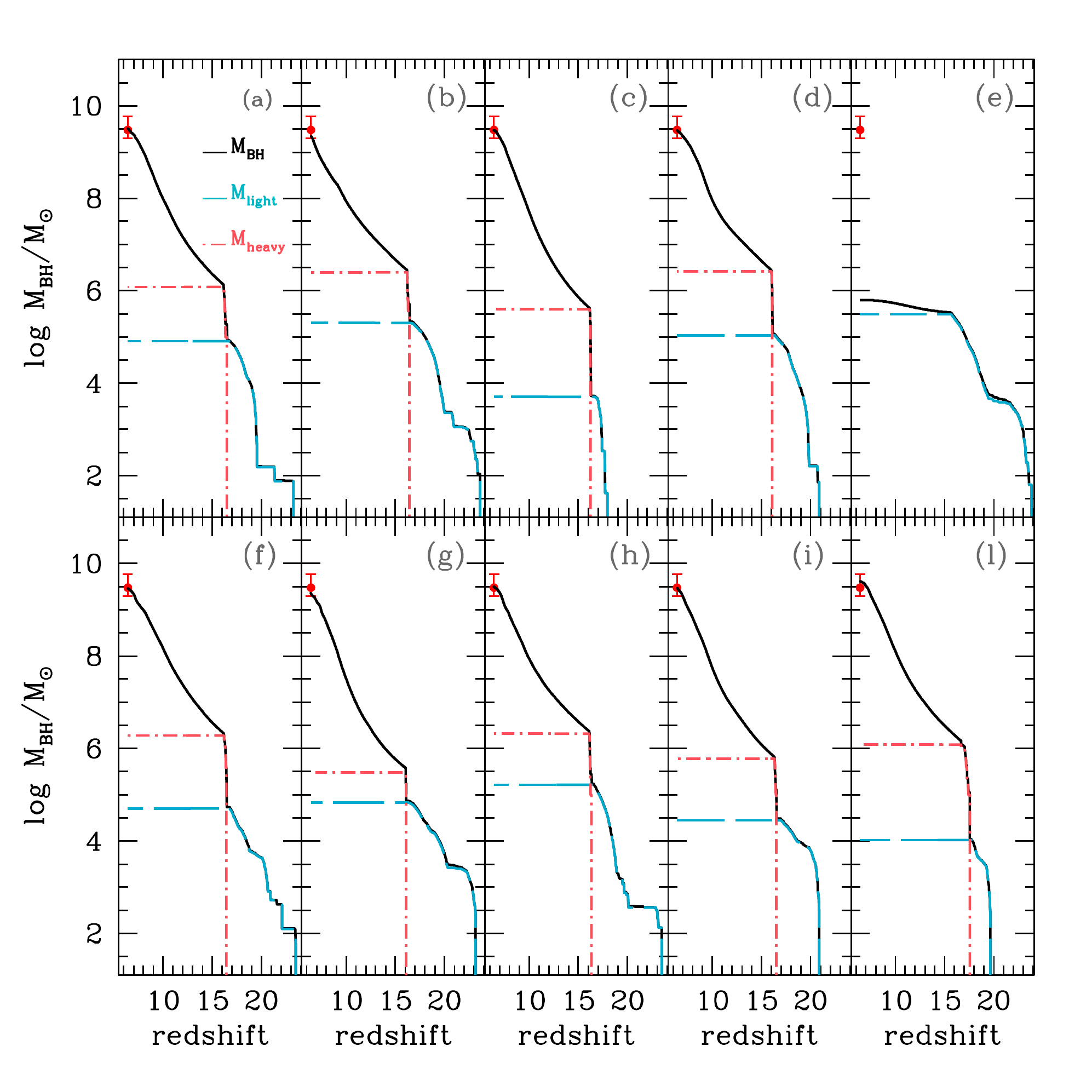}
\caption{Evolution of the BH mass as a function of redshift for 10 different merger trees. Black solid lines show the total BH mass while blue dashed and red dot-dashed
lines represent the contribution of light and heavy seeds, respectively, without gas 
accretion.} 
\label{fig:bhevo10r} 
\end{figure} 

\subsection{Dependence on the hierarchical history}
\label{sec:singelMTs}

One of the advantages of a semi-analytical model is that it allows one
to run independent merger tree simulations of the same quasar.

In Fig.~\ref{fig:bhevo10r} we show the evolution of the BH mass and
the contribution of light and heavy seeds as a function of redshift for
10 different merger trees. In 9 out of 10 runs, the final BH mass is in 
very good agreement with the data. 
The only exception is the simulation shown in panel (e),
where only light BH seeds form. This supports the conclusion that,
 as long as gas accretion is assumed to be Eddington-limited,
heavy BH seeds are required to grow a SMBH at $z > 6$.

The relative contribution of light and heavy BH seeds depends on the individual
merger tree. 
Even when light BH seeds start to form at
$z \gtrsim 20$ (see
panels a, b, f, g and h), their total mass does not exceed $\sim 10^5\, \rm M_\odot$ and it is comparable to
the mass of one single heavy BH seed.  Only
between $\sim 3$ to $\sim 30$ heavy BH seeds are required to grow a SMBH by $z \sim 6.4$, 
and their total mass ranges between $\sim 3 \times 10^5 \, \rm M_\odot$ and 
$\sim 3 \times 10^6 \, \rm M_\odot$.

\begin{figure}
\vspace{\baselineskip}
\includegraphics[width=\hsize]{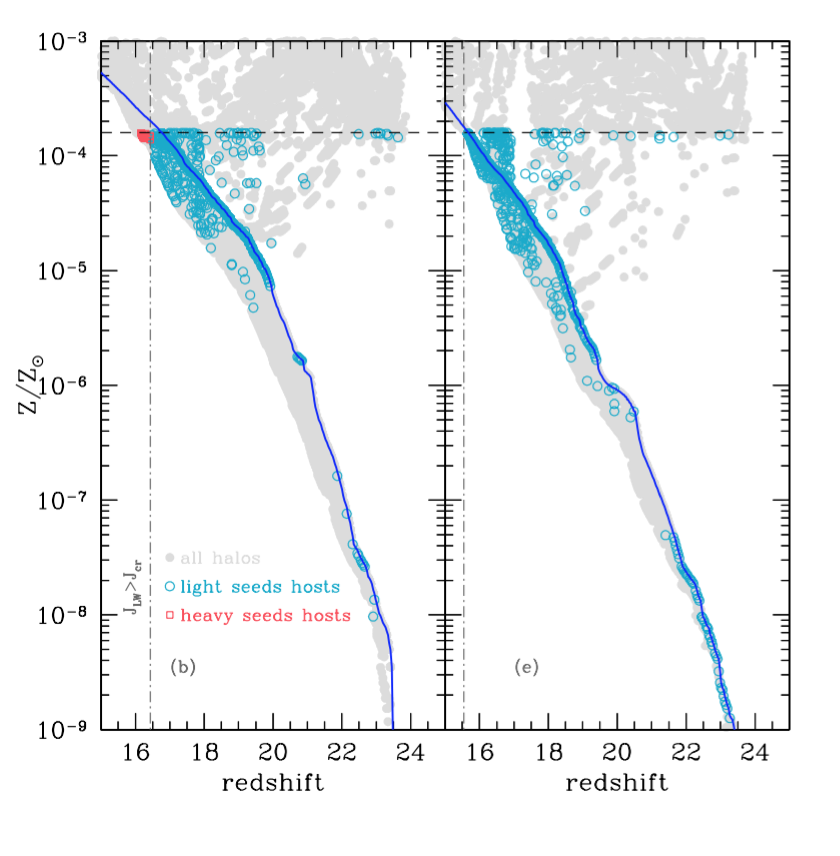}
\caption{Metallicity as a function of redshift of all the progenitor halos (gray points) in two merger
tree simulations (b, left panel) and (e, left panel) shown in Fig.~\ref{fig:bhevo10r}. Open blue circles
and red squares indicate the progenitors where light and heavy BH seeds form. In each panel,
the solid line is the mean IGM metallicity, the horizontal dashed line is the value of $Z_{\rm cr}$
and the vertical dot-dashed line is the redshift at which $J_{\rm LW} > J_{\rm cr}$.} 
\label{fig:agemet} 
\end{figure}

It is interesting to investigate in more details why no heavy BH seed is formed
in the simulation shown in panel (e). We compare the properties of this simulation
with the one shown in panel (b), which is characterized by a similar high-$z$ evolution
of the light BH seeds mass. In Fig.~\ref{fig:agemet} we show the metallicity of 
all progenitor halos in the two simulations (gray points) as a function of redshift.
The solid line is the mean metallicity of the IGM and open blue circles 
(red squares) represent progenitor halos hosting light (heavy) BH seeds.
Light BH seeds form in halos with $Z < Z_{\rm cr}$, where the value corresponding
to $Z_{\rm cr}$ in the reference model is shown by the horizontal dashed line. 
As expected, heavy seeds form only  in a small number of halos of simulation (b), 
where $Z < Z_{\rm cr}$ and $J_{\rm LW} > J_{\rm cr}$ 
(the redshift at which this condition is satisfied is indicated by the vertical 
dot-dashed line). 

The redshift and metallicity distribution of progenitor halos is different in the two
simulations. Newly virialized progenitor halos  have the same metallicity of the IGM, while 
in others the metallicity can be significantly smaller or larger.  
At $z \lesssim 20$, $Z_{\rm IGM}$ is slightly smaller 
in simulation (e) and there is a smaller fraction of halos with $Z < Z_{\rm IGM}$,
meaning that self-enrichment is more efficient than in simulation (b).
In addition, the LW background intensity becomes larger than $J_{\rm cr}$ at a lower
redshift in simulation (e),  $z \sim 15.5$ instead of $z \sim 16.5$.  
As a result, there is no single progenitor halo where  the conditions for heavy BH seed formation
are satisfied.

\begin{figure*}
\hspace{-1 cm}
\includegraphics [width=6.0cm]{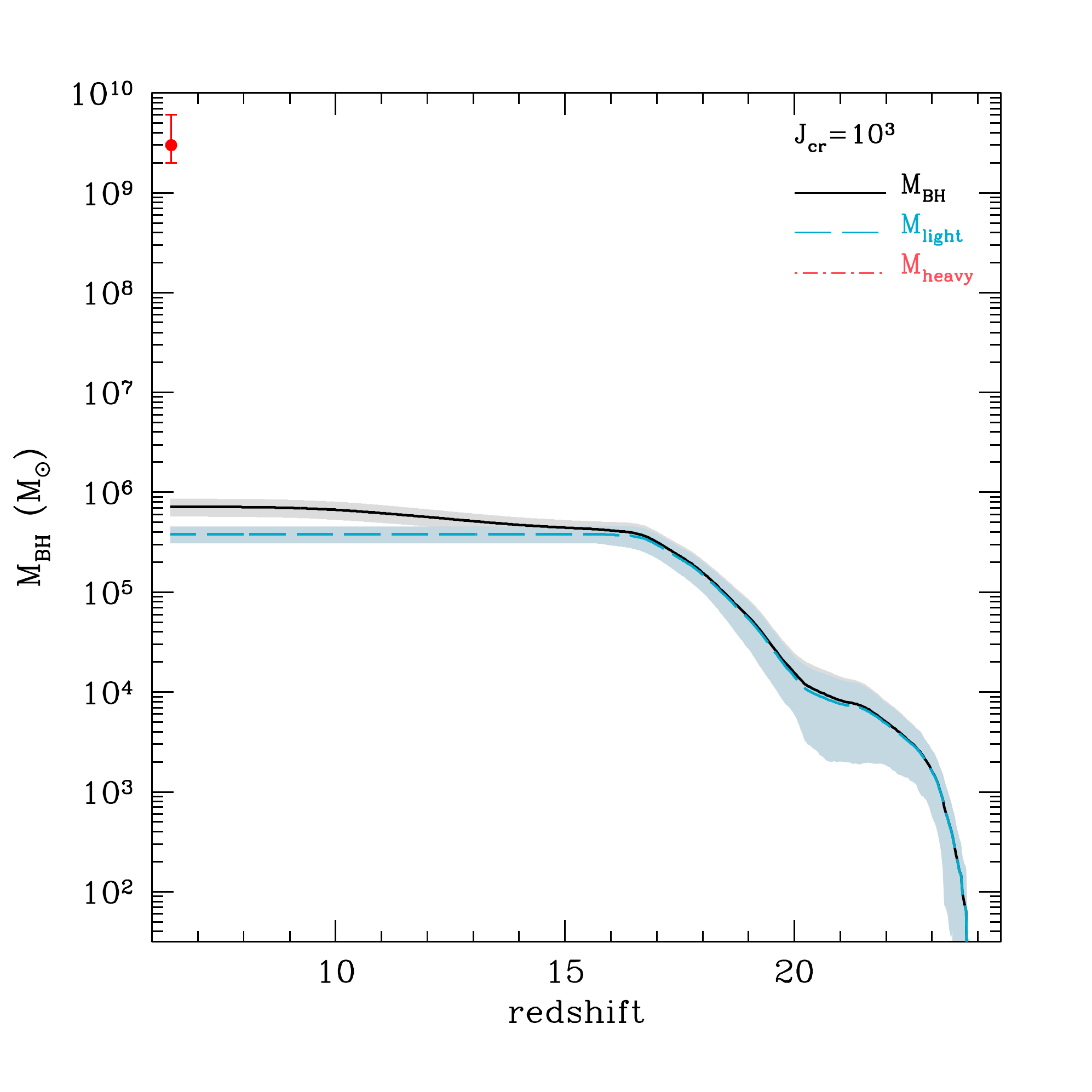}
\includegraphics [width=6.0cm]{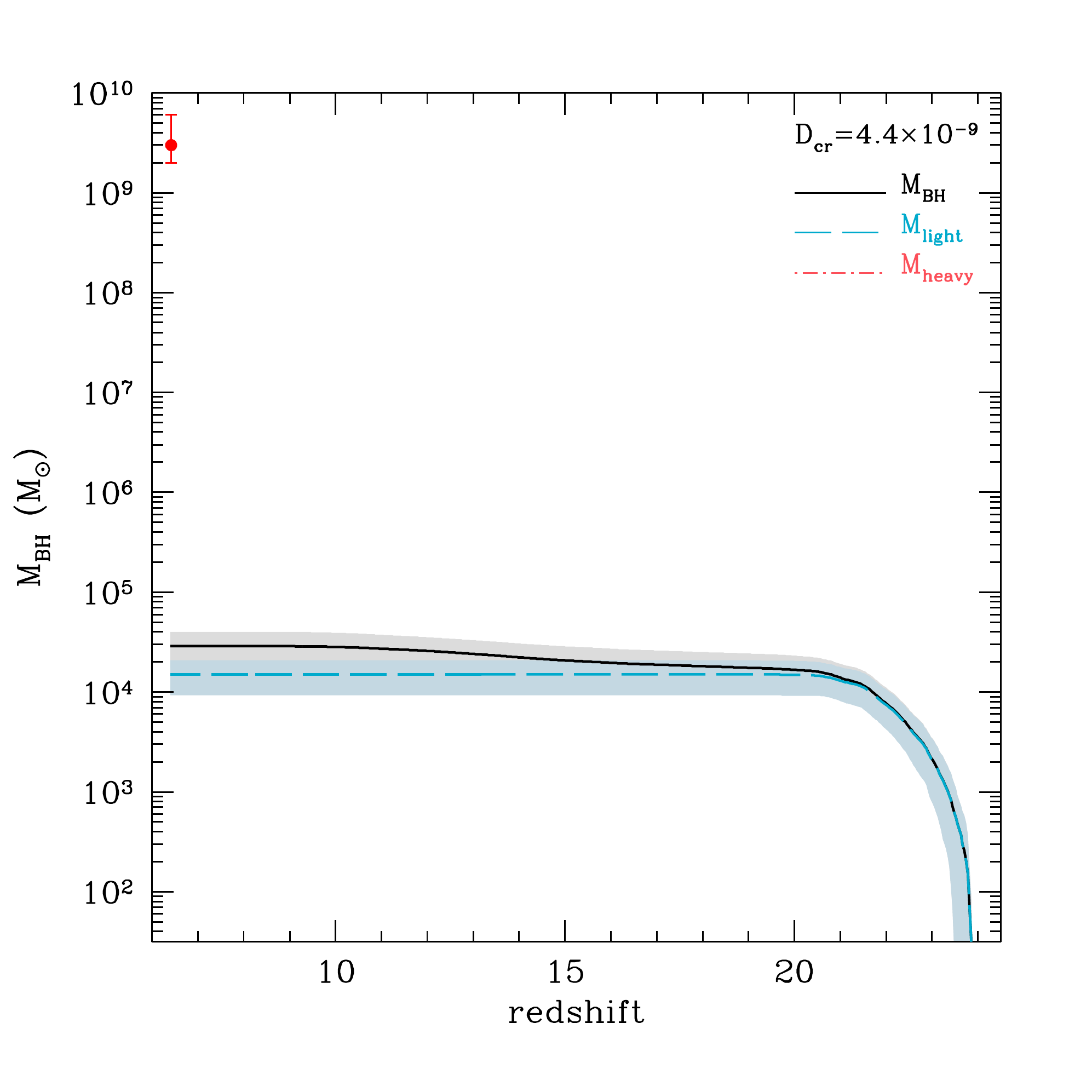}
\includegraphics [width=6.0cm]{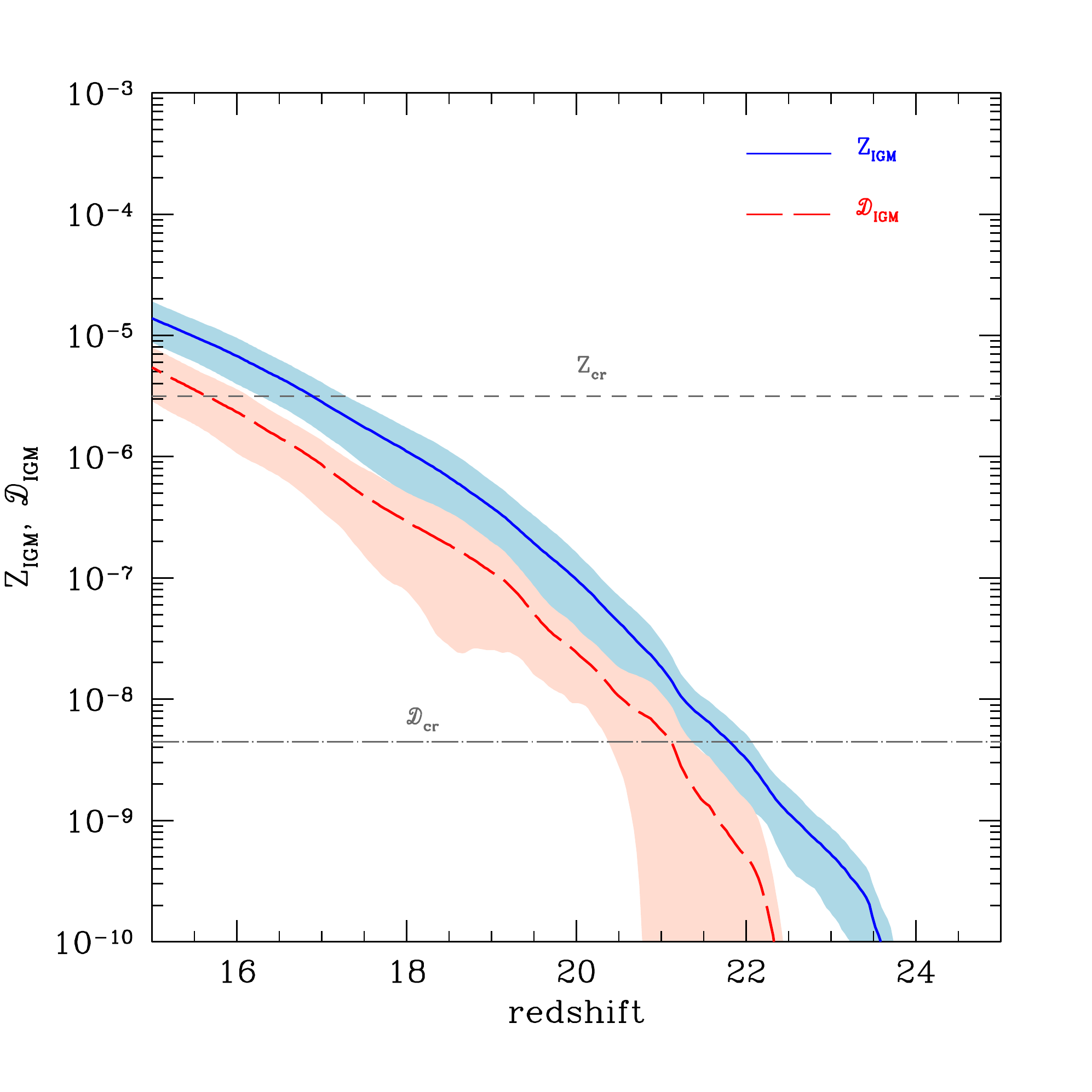}
\caption{Same as in Fig.~\ref{fig:bhevo} but assuming  $J_{\rm cr} = 10^3$ (left panel) and 
a dust-driven transition to Pop~II star formation occurring at ${\cal D}_{\rm cr} = 4.4 \times 10^{-9}$ (middle panel). 
In the right panel we show the average metallicity (blue solid line) and dust-to-gas ratio (red dashed 
line) of the IGM. Horizontal dashed (dot-dashed) line indicates $Z_{\rm cr}$ in absolute units and ${\cal D}_{\rm cr}$, respectively. 
In all panels the curves are averages over 10 merger tree realizations with shades representing the $1-\sigma$ dispersion.} 
\label{fig:testLWDcr} 
\end{figure*} 

\section{Discussion} 
\label{sec:discussion}

The results of the reference model depend on a number of assumptions whose
importance is critically discussed below.

\subsection{Dependence on $J_{\rm cr}$}

The critical intensity of the LW background that enables the collapse of gas in 
metal-poor Ly$\alpha$ halos is still highly debated. Here
we discuss the consequences of increasing $J_{\rm cr}$ to  $10^3$.

Although the LW background can reach very large values in the biased region that we are
simulating, on average it exceeds $J_{\rm cr}=10^3$ at $z \lesssim 14.5$ (see left panel in Fig.~\ref{fig:feedback}),
when all progenitor halos have been already enriched above the critical metallicity for Pop~II star formation.
Ly$\alpha$ halos with $Z < Z_{\rm cr}$ and 
$300 \lesssim J_{\rm LW} \lesssim 10^3$ now host Pop~III star formation and no single heavy BH seed forms.

The redshift evolution of the total BH mass is shown in the left panel of Fig.~\ref{fig:testLWDcr}. 
The  growth of the BH is strongly suppressed and $M_{\rm BH} \lesssim 10^6 \, \rm M_\odot$ at  
$z \sim 6.4$. In fact, despite the larger number of light seeds at 
$16 \lesssim z \lesssim 18$ compared to the reference model ($\sim 4$ times larger, on average), 
their BH masses are too small to activate efficient 
gas accretion, unless a much higher BH accretion efficiency ($\alpha_{\rm BH}$) or super-Eddington accretion
is assumed (\citealt{VolonteriRees05, Li12, AN14, MHD14}; Pezzulli et al. 2015).

\subsection{Dependence on the critical dust-to-gas ratio}

In the reference model we assume that low mass Pop II stars form
when the metallicity of the star forming gas reaches a critical value of $Z_{\rm cr} = 10^{-3.8}$
Z$_\odot$,
above  which metal fine-structure line cooling becomes efficient. 
However, semi-analytic and numerical studies suggest that gas cooling and fragmentation
can be activated at a lower metallicity when dust grains are present (\citealt{Schneider02}; \citealt{Schneider06};
 \citealt{Omukai10}; \citealt{Dopcke11}). 
\citet{Schneider12a} show that low-mass Pop~II stars can form
when the dust-to-gas mass ratio, ${\cal D}$, exceeds a critical value of
${\cal D}_{\rm cr}=4.4^{-1.9}_{-1.8} \times 10^{-9}$. 
Moreover, a dust-driven transition is consistent with observations of the
tail of the metallicity distribution function of Galactic halo stars \citep{Schneider12b,deBen14}.

In \gamete \, we follow dust enrichment in the ISM of all progenitor halos and we can explore
the effects of dust cooling and fragmentation on the formation of Pop~III
stars, hence of light BH seeds, and on the direct collapse of gas onto a heavy BH seed. 
Following \citet{Omukai08}, we assume that when ${\cal D} < {\cal D}_{\rm cr}$
and $J_{\rm LW} > J_{\rm cr}$, the gas collapses almost isothermally until the densities are
large enough to activate dust cooling and fragmentatio, forming a compact 
Pop~II stellar cluster. The middle panel of Fig.~\ref{fig:testLWDcr} shows
that the effect is similar to imposing a larger LW flux critical threshold. A smaller number of
light seeds is formed, heavy seed formation is suppressed, and BH growth is dramatically inefficient, 
leading to $M_{\rm BH} \sim 3 \times 10^4 \, \rm M_\odot$ at $z=6.4$. In fact, on average, dust enrichment
allows most of the halos to reach the critical threshold at $z \gtrsim 20$ (see the right panel
of Fig.~\ref{fig:testLWDcr}), confining the formation
of light seeds only in the first star forming progenitors and preventing the formation of heavy seeds,
as the condition $J_{\rm LW} > J_{\rm cr}$ is achieved only at smaller redshifts. 

This conclusion
does not depend on the fate  of the newly formed dust-induced compact Pop~II stellar clusters. Even assuming that
their dynamical evolution favors the collapse into a black hole of mass $\sim 10^3 \, \rm M_\odot$ 
\citep{Omukai08,Devecchi09, Devecchi10, Devecchi12}, their number and mass are too small to significantly affect the BH mass growth rate.

\begin{figure}
\includegraphics [width=7.5cm]{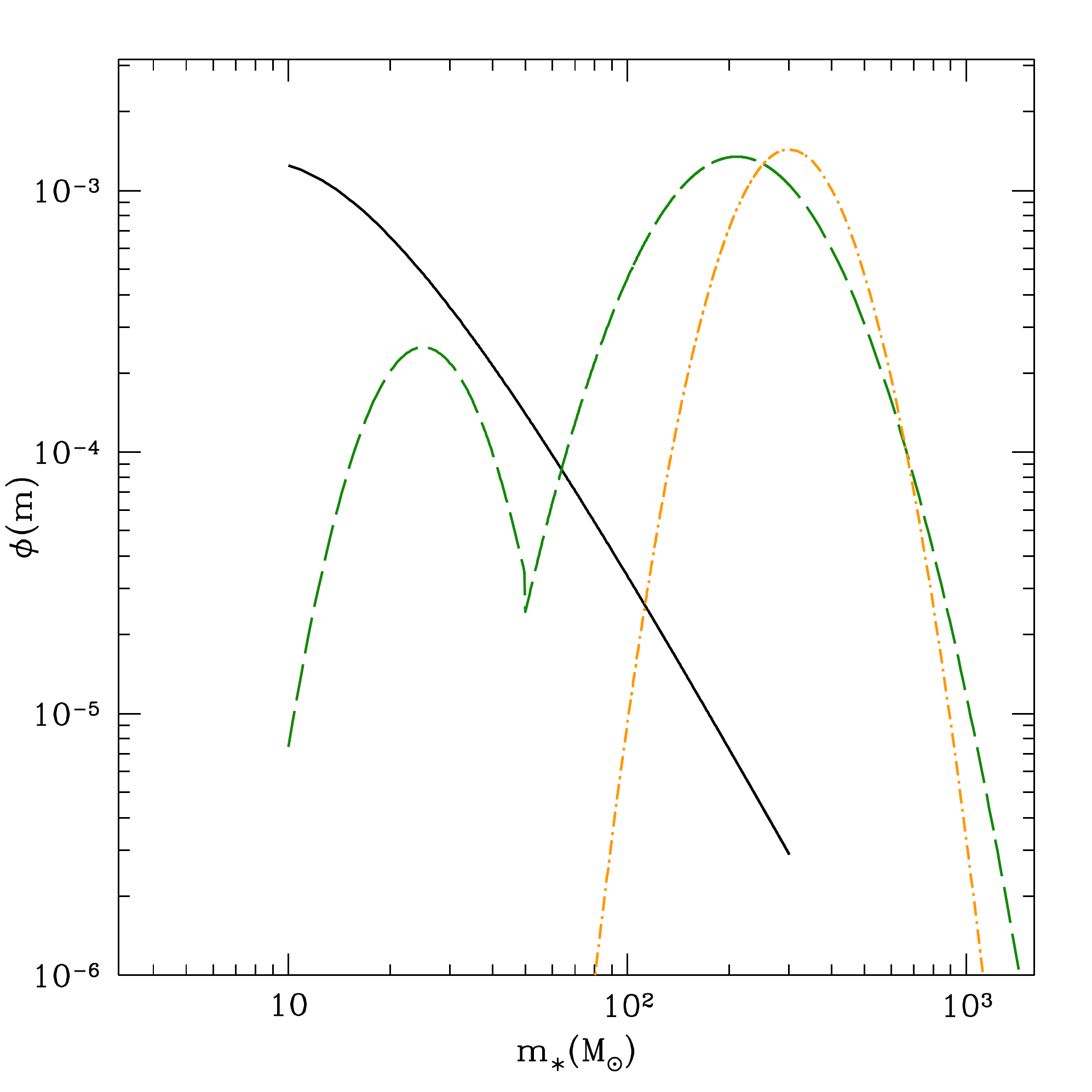}
\caption{Initial mass function (IMF) of Population III stars. The black solid line shows the Larson IMF
adopted in the reference model, normalized to 1 in the mass range $[10-300]\, \rm M_\odot$. 
Green dashed  and orange dot-dashed lines show the analytic functions used to approximate
the results of \citet{Hirano15} for $J_{\rm LW} < 0.1$ and $>0.1$, respectively. 
In both cases, the mass distribution is normalized to 1 in the mass range $[10-2000]\, \rm M_\odot$.} 
\label{fig:hiranoIMF} 
\end{figure} 
\begin{figure*}
\hspace{-1 cm}
\includegraphics [width=6.0cm]{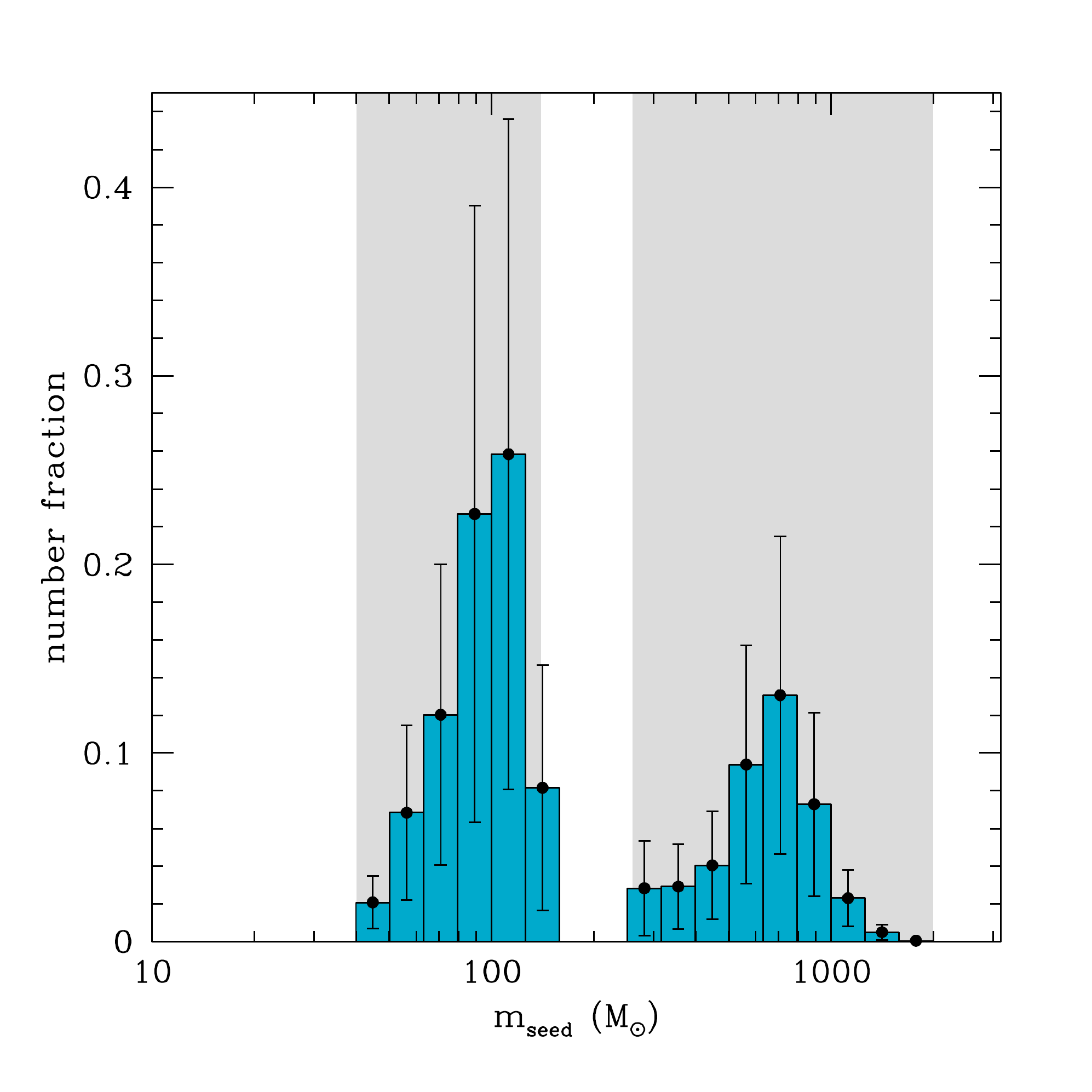}
\includegraphics [width=6.0cm]{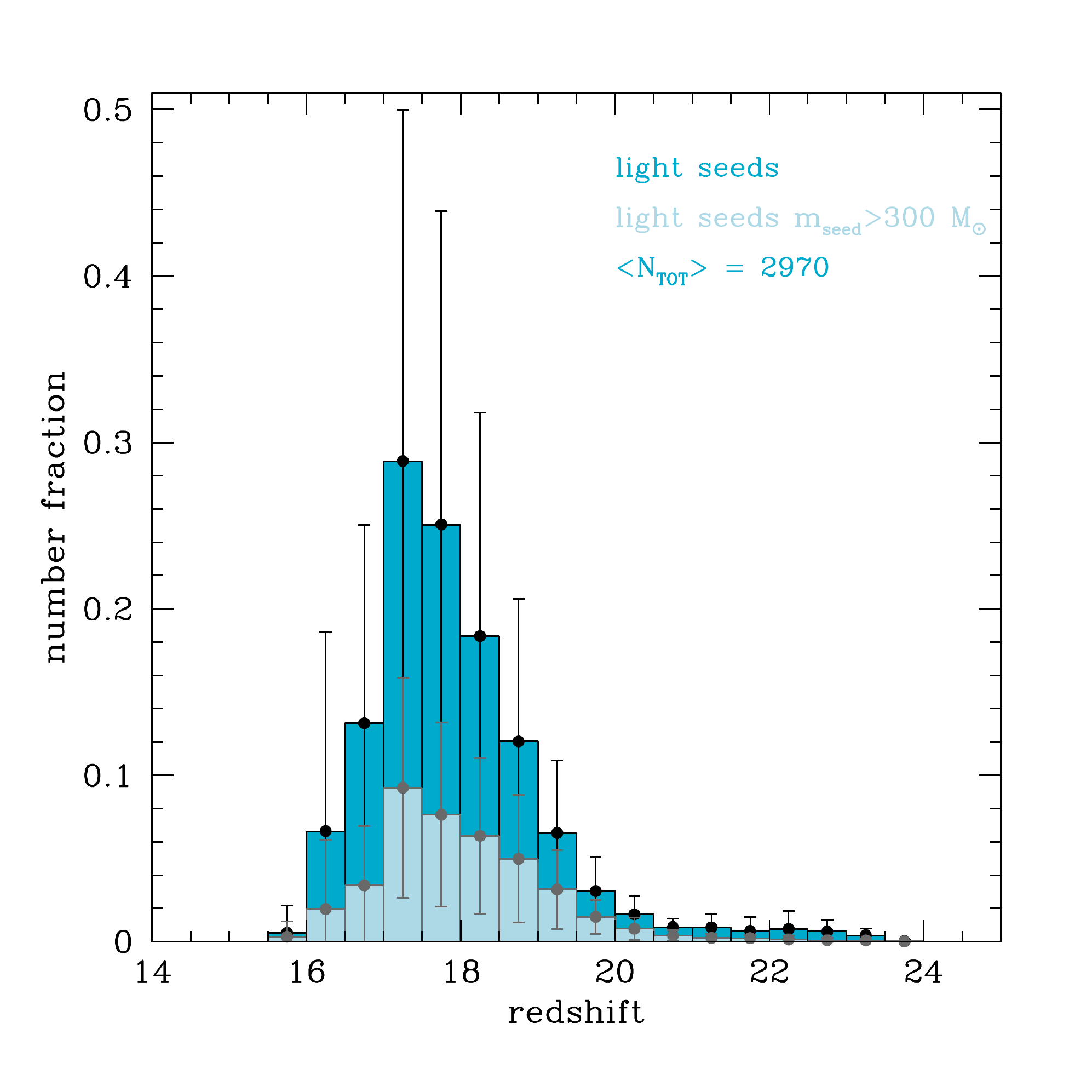}
\includegraphics [width=6.0cm]{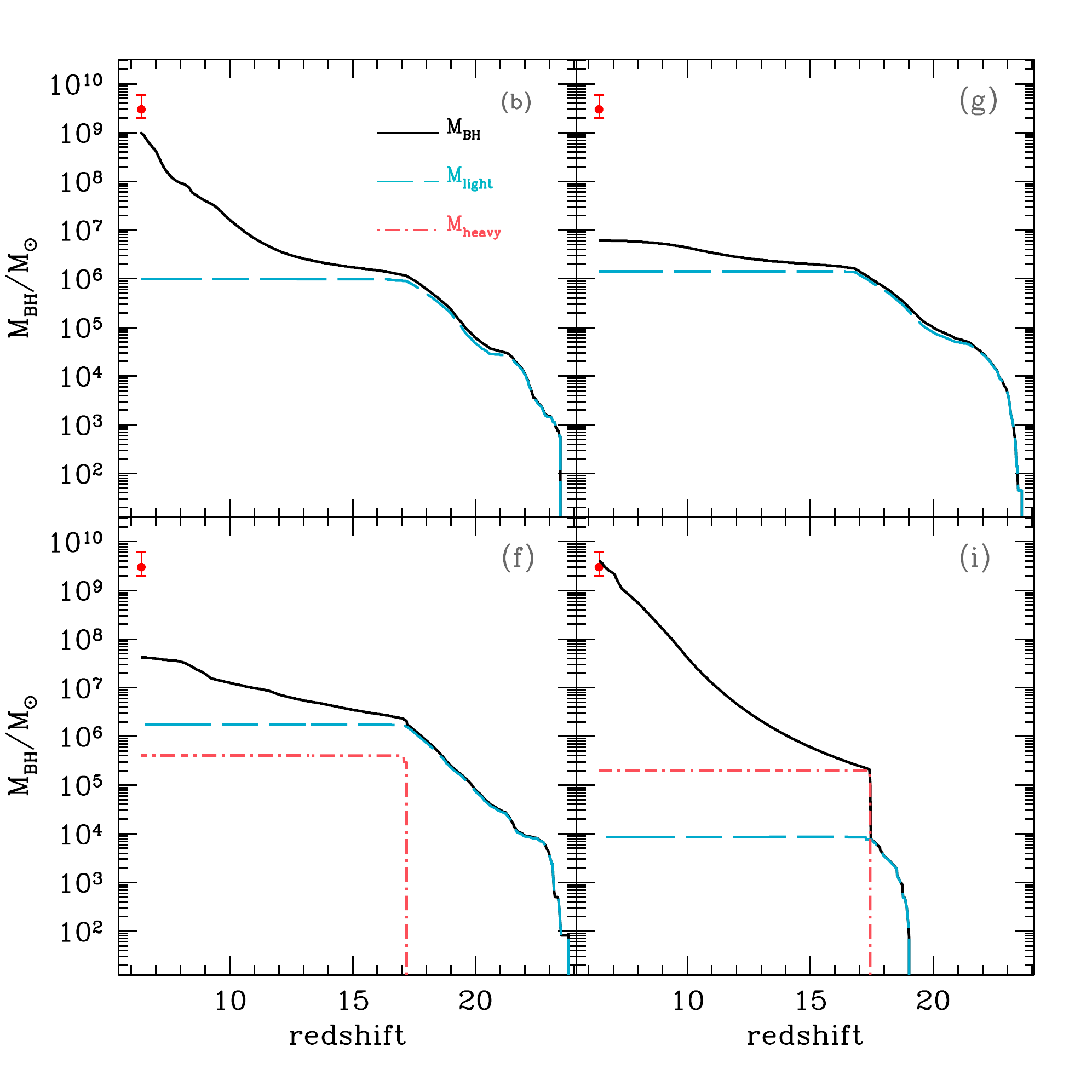}
\caption{Same as Fig.~\ref{fig:histogLS} but only for light BH seeds and 
assuming the environment-dependent
Pop~III IMF predicted by \citet{Hirano15}. In the right panel we show the
BH mass growth and the contribution of light and heavy seeds in 4 different realizations
of the merger tree. The labels (b), (g), (f) and (i) are used to enable a direct comparison 
with the results presented in Fig.~\ref{fig:bhevo10r}.} 
\label{fig:testhirano} 
\end{figure*} 

\subsection{IMF of Population III stars}

First attempts to predict the Pop III stellar mass spectrum ab-initio, starting from cosmological
initial conditions, have been recently made through sophisticated numerical simulations
\citep{Hirano14, Hirano15}. 
According to \citet{Hirano15}, Pop III stars which form in halos exposed to a LW background intensity $J_{\rm LW}<0.1$ 
follow a mass distribution characterized by two peaks, at $\sim 25$ and 
$\sim 250 \, \rm M_\odot$.
Conversely, only very massive ($>100 \, \rm M_\odot$) stars form when $J_{\rm LW}>0.1$ 
 as less efficient cooling causes higher gas temperature and 
larger accretion rates (see Fig.~6 in \citealt{Hirano15}). 

In order to test the implications of this environment-dependent Pop~III IMF, we 
approximate the mass distribution found by \citet{Hirano15} with the analytic
functions shown in Fig.~\ref{fig:hiranoIMF}, where the IMF are normalized
to 1 in the stellar mass range $[10-2000]\, \rm M_\odot$ (dashed and dot-dashed lines
for $J_{\rm LW} < 0.1$ and $>0.1$, respectively). For comparison, we
also show the IMF adopted in the reference model (solid line). 
Given the shape of the new distribution, we expect a larger number
of massive Pop III remnants, leading to more frequent light BH seeds with 
mass $> 300 \, \rm M_\odot$ compared to the reference model.

The resulting average mass and redshift distributions of light seeds are
shown in the left and middle panels of Fig.~\ref{fig:testhirano}. The two
peaks of the mass distribution reflect the underlying Pop~III IMF. 
The number of light BH seeds is $\sim 10$ times larger than in the
reference model, and $\sim 30 \%$ of these have a mass $> 300 \, \rm M_\odot$. 
In fact, the shape of the underlying Pop~III IMF affects the star formation
history at $z > 15$ through both mechanical and chemical feedback: the larger
number of stars with masses in the pair-instability SN range,
$160 \, \rm M_\odot \lesssim m_\ast \lesssim 240 \, \rm M_\odot$ \citep{Heger02}, and 
above leads to strong SN and AGN-driven outflows of metal-enriched gas out of the first mini-halos. 
The integrated effect of this feedback-regulated star formation rate along the
hierarchical evolution is to decrease the metallicity of gas-poor star forming
progenitors - favoring the formation of a larger number of Pop~III BH remnants -
and the LW emissivity, hence the intensity of the LW background. As a result,
the condition $J_{\rm LW} > J_{\rm cr}$ is met at lower $z$ compared to the
reference model, when most of the Ly$\alpha$ halos have already been enriched
above the critical metallicity and  the formation of heavy seeds
is suppressed in 7 out of 10 merger trees.

In these conditions, the BH mass growth at $10 \lesssim  z \lesssim 15$ is very sensitive to the amount 
of leftover gas from winds in progenitor systems. In the right panel of  Fig.~\ref{fig:testhirano} we show the 
BH mass growth and the contribution of light and heavy seeds as a function of redshift for 4 different merger tree 
simulations. In the top panels, we show the results for the same (b) and (g) 
merger tree realizations presented in Fig.~\ref{fig:bhevo10r}.  Despite 
no heavy seed is formed, $M_{\rm BH} \sim 10^9 \, \rm M_\odot$ at $z \sim 6.4$
in simulation (b), only a factor of a few smaller than the observed value.
Although the total BH mass contributed by light seeds is similar, gas accretion
is less efficient in simulation (g) and the final BH mass is significantly smaller. 
In the bottom panel, we show the results of two simulations where, despite a 
comparable number of heavy seeds forms,  the
final BH mass at $z \sim 6.4$ differs by almost two orders of magnitude.
In simulation (f), a large number of light BH seeds forms over the redshift range
$17 \lesssim z \lesssim 25$. Gas depletion in their progenitor galaxies
due to AGN feedback causes a lower average accretion rate at $z < 15$ and
the BH mass at $z = 6.4$ is only $\sim 5 \times 10^7 \, \rm M_\odot$. On the contrary,
the small number of light BH seeds formed in simulation (i) at $z \lesssim 20$
does not significantly affect the gas content of progenitor galaxies and gas accretion
at $z < 15$ can be efficient enough to form a SMBH at $z = 6.4$, with a mass consistent
with the data.

\begin{figure*}
\hspace{-1 cm}
\includegraphics [width=6.0cm]{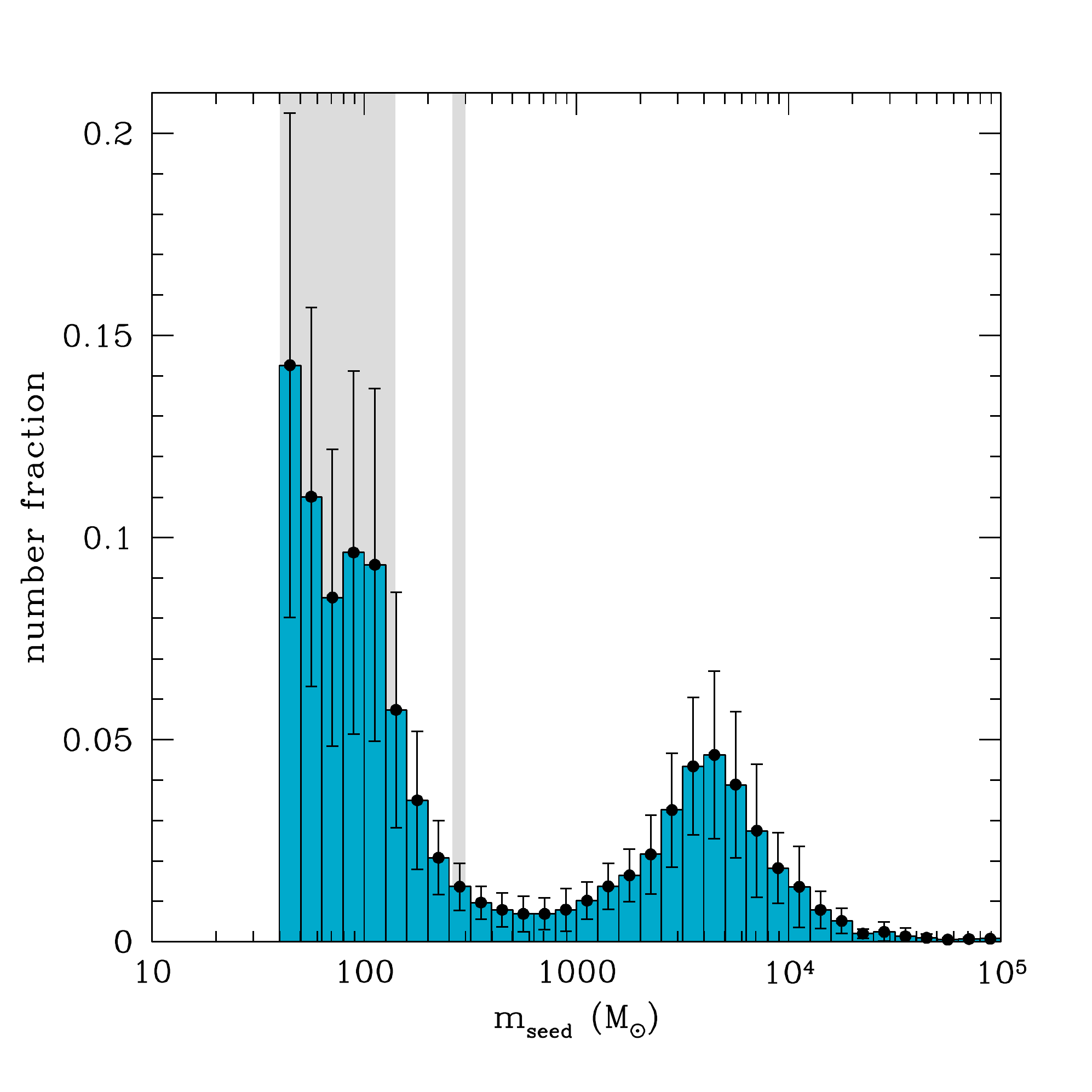}
\includegraphics [width=6.0cm]{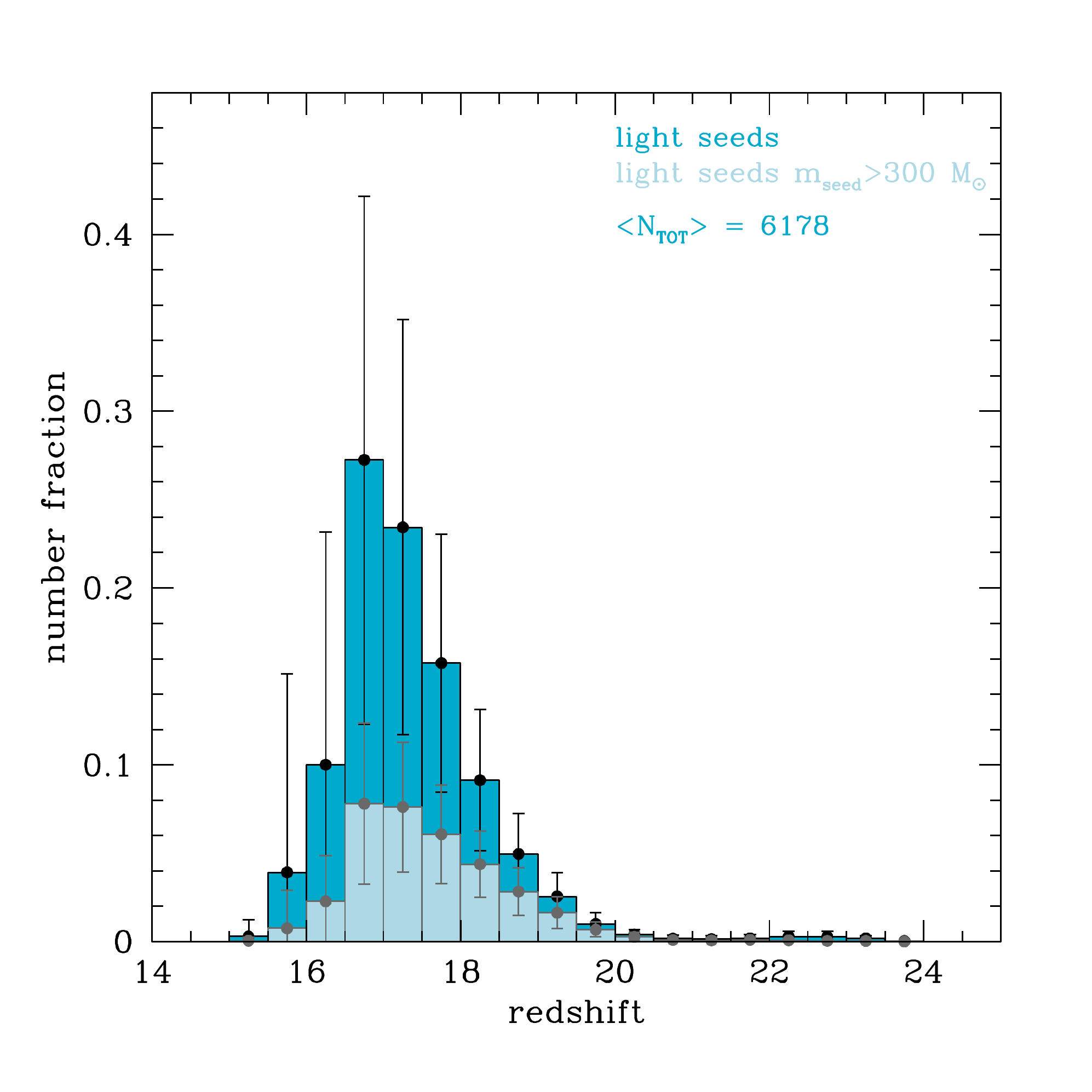}
\includegraphics [width=6.0cm]{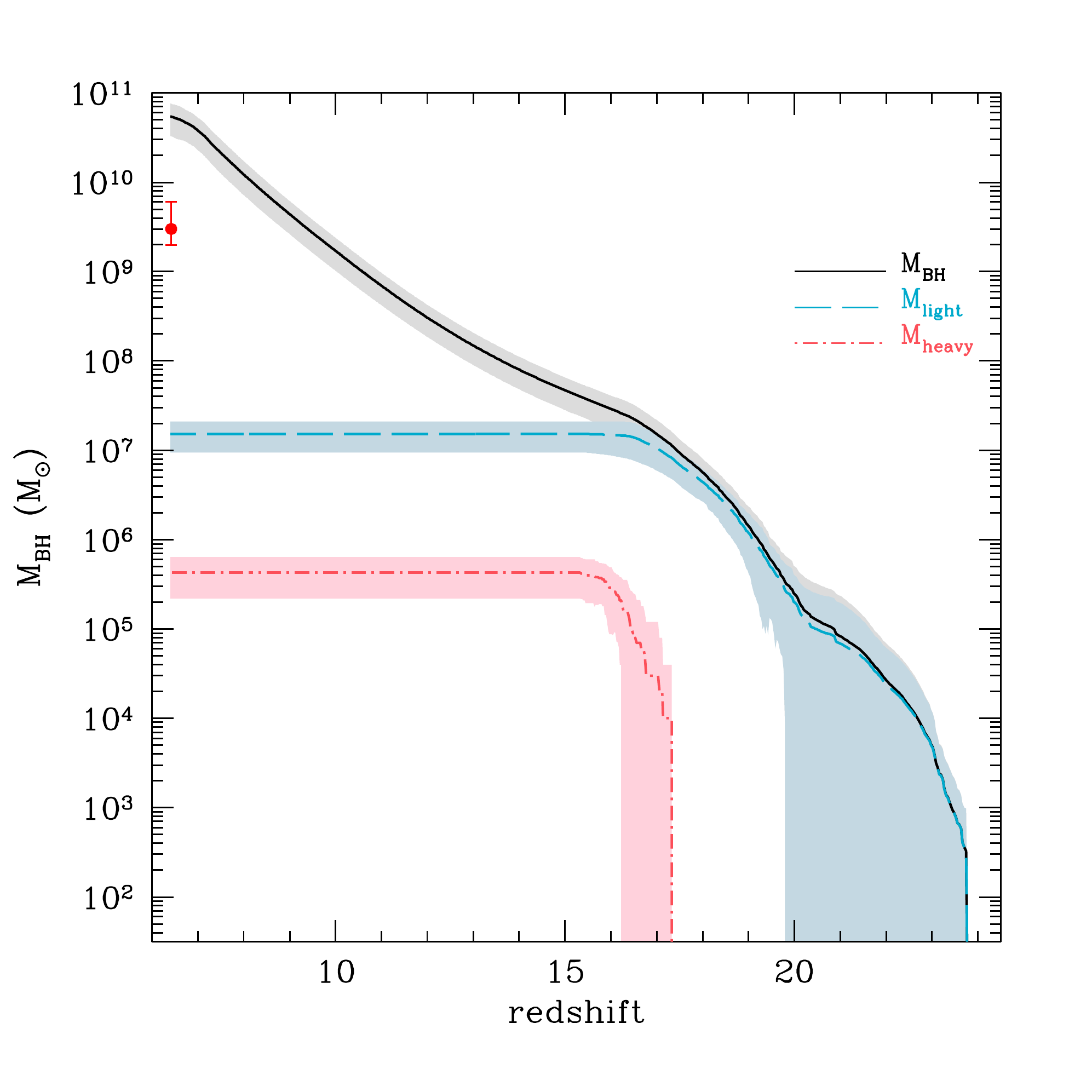}
\caption{Same as in Figs.~\ref{fig:bhevo} and~\ref{fig:histogLS} but assuming that
all BH remnants of Pop III stars merge to form a single more massive BH seed.
In addition, in the middle panel we show the redshift distribution of light seeds with
 masses $>300$ \msun \, (azure histogram and gray data points).} 
\label{fig:testdynamics} 
\end{figure*} 
\subsection{Dynamics of light BH seeds}

In the reference model, we assume that only the most massive  Pop III remnant form a light BH seed that
settles at the galaxy center accreting gas 
from the surrounding medium. 
To test the effect of this assumption, here we investigate the opposite, extreme scenario
and we allow all BH remnants to merge and form a single more massive 
light BH seed that migrates at the center of the galaxy. The underlying 
assumption is that the merging timescale is shorter than the characteristic timestep
of the simulation, and that dynamical effects, such as 3-body scattering \citep{Miller02, Gultekin06} and gravitational recoil 
\citep{Haiman04, Shapiro05, VolonteriRees06} are not ejecting the merging BHs. Regarding 3-body scattering, 
we can rescale the results by \citeauthor{Gultekin04} (\citeyear{Gultekin04, Gultekin06}) and \cite{Miller09}, while for the gravitational 
recoil we can use the Monte Carlo sampling of recoil velocities in \cite{VGD10}. In general, the potential well of the halo
becomes sufficiently massive to retain merging and scattering BHs only when the halo mass is $\sim 10^7-10^8 \, \rm M_\odot$. This is because most
of the merging BHs have mass ratio between 1:6 and 1:1, where the recoil velocity is typically $>100$ km/s for random spin magnitudes and configurations. Similarly, 3-body scattering appears to cease to be effective in ejecting BHs when the escape velocity becomes $>100$ km/s. 
We note, however, that sudden gas inflows, triggered by mergers or collimated gas streams from the cosmic web, can temporarily deepen the potential well
and allow mergers and 3-body scattering to occur \citep{Davies11,Lupi14}. We will in the following assume, optimistically, that all formed BHs can merge, but in reality we expect that only a fraction of them can be retained, and this fraction increases with the mass of the host halo.

The resulting mass and redshift distribution of light seeds is shown in the left and middle 
panels of Fig.~\ref{fig:testdynamics}, respectively. As expected, the mass spectrum of
light seeds now extends well beyond $300 \, \rm M_\odot$, the maximum BH
remnant mass for the adopted Pop~III IMF. Less massive, more numerous, light seeds
($\lesssim 300 \, \rm M_\odot$) form in less efficient star 
forming halos, while more massive light seeds ($> 300 \, \rm M_\odot$) are the result of the 
coalescence of several (from few to hundreds) Pop III remnants formed in 
more efficient star forming halos. Interestingly, there is a tail of the mass distribution
that extends up to $\sim 10^5 \, \rm M_\odot$, showing that few light seeds may reach a
mass comparable to that of heavy seeds. 

The redshift distribution of light seeds is fairly independent of their mass, as shown
by the two histograms  in the middle panel of Fig.~\ref{fig:testdynamics}.
Compared with the analogous plot for the reference model shown in Fig.~\ref{fig:histogLS},
a much larger number of light seeds is formed. This is a consequence of the
stronger feedback induced by more massive BHs on their host galaxy. In the shallow
potential wells of small halos at high redshift, BH feedback is able to unbind most 
- if not all - of the gas. As a result, the ISM metallicity remains below the critical value for 
a longer period of time, leading to a prolonged phase of Pop~III star and light BH seeds
formation. 

The effect on the BH mass growth rate is shown in the right panel of Fig.~\ref{fig:testdynamics}.
In this case, the contribution of light seeds exceeds that of heavy seeds, which is smaller
(by a factor $\sim 3$) than in the reference model, and triggers a faster and more efficient growth.
The BH mass exceeds $\sim 10^9 \, \rm M_\odot$ at $z \lesssim 10$ and reaches a final value
of  $\sim 6 \times 10^{10} \, \rm M_\odot$ at $z \sim 6.4$, a factor of 20 larger than in the reference model.

\section{Conclusions} 
\label {sec:conclusions}

We have investigated the origin of SMBHs at $z > 6$ applying a
largely improved version of the semi-analytical model \gamete. 
In this work we explore the relative role of light BH seeds, formed as remnants
of massive Pop~III stars, and heavy BH seeds, formed by the direct collapse
of gas, in the formation pathway to the first SMBHs. 

To this aim, we have implemented a physically motivated prescription to estimate
the cold gas mass fraction in mini-halos, taking into account molecular and metal
fine-structure cooling and the photo-dissociation of $\rm H_2$ in the presence of an
external LW background. 
We then follow the subsequent
evolution of the BHs and their host galaxies along the hierarchical history of a
$z = 6.4$ halo with a mass of $10^{13} \, \rm M_\odot$.
The free parameters of the model, such
as the accretion efficiency entering in the formulation of Eddington-limited
Bondi accretion, the AGN wind efficiency, the efficiency of quiescent and merger-driven
star formation, have been fixed to reproduce the observed properties of SDSS J1148.
Simulating different merger trees of the same halo, we compute the intensity of the
LW background, accounting for the contribution of stars and accreting BHs, the
filling factor of ionized regions, the metal and dust enrichment in and outside the
progenitor galaxies, to explore if and when heavy BH seeds can form in metal-poor
Lyman-$\alpha$ halos exposed to a strong LW background. 

In the reference model, where we assume that Pop~III stars form in progenitor
galaxies with $Z < Z_{\rm cr} = 10^{-3.8}$ Z$_\odot$ according to a Larson IMF in the 
mass range $10 \, \rm M_\odot \le m_\ast \le 300 \, \rm M_\odot$, a small number of light
BH seeds are hosted in mini-halos at $z \gtrsim 20$ before radiative feedback is
able to suppress $\rm H_2$ cooling. The dominant fraction of light BH seeds form in 
Lyman-$\alpha$ cooling halos at $15 \lesssim z \lesssim 20$, before the intensity
of the LW background becomes larger than $J_{\rm cr} = 300$ allowing the direct
collapse of gas and the formation of heavy seeds. In these conditions, we find that
in 9 out of 10 merger tree simulations between 3 to $\sim 30$ heavy seeds are able to form 
before metals have enriched all the progenitor galaxies  to $Z \ge Z_{\rm cr}$ and 
low-mass Pop~II stars form. We find that:

\begin{itemize}

\item The growth of $z > 6$ SMBHs relies
on heavy seeds. The only simulation where the interplay between chemical and 
radiative feedback effects prevents the formation of heavy seeds predicts a
final BH mass of $M_{\rm BH} < 10^6 \, \rm M_\odot$.

\item The above result dramatically depends on the assumed values of $J_{\rm cr}$
and $Z_{\rm cr}$. A larger value of $J_{\rm cr}$ ($= 10^3$) or a Pop~III/Pop~II transition
driven by dust-cooling at a critical dust-to-gas ratio of ${\cal D}_{\rm cr} = 4.4 \times 10^{-9}$
prevents the formation of heavy seeds, hampering the mass growth of the nuclear BH
so that its final mass at $z = 6.4$ is $M_{\rm BH} < 10^6 \, \rm M_\odot$ in all the merger tree
simulations.

\item The relative importance of heavy and light BH seeds depends on the adopted IMF
of Pop~III stars, as this affects the history of cold gas along the merger tree by means of
SN and AGN-driven winds.

\item As long as gas accretion is assumed to be Eddington-limited, the mass of individual BH 
seeds is the key condition to trigger SMBH growth. If  all BH remnants merge before 
settling to the center of their progenitor galaxy, the mass distribution of light BH seeds 
extends to $\sim 10^3 - 10^5 \, \rm M_\odot$. In these conditions, gas accretion is so efficient 
that by $z = 6.4$ the SMBH mass is - on average - $ >10^{10} \, \rm M_\odot$ and the evolution is 
completely dominated by light BH seeds. 
\end{itemize}

\begin{figure}
\includegraphics [width=7.5cm]{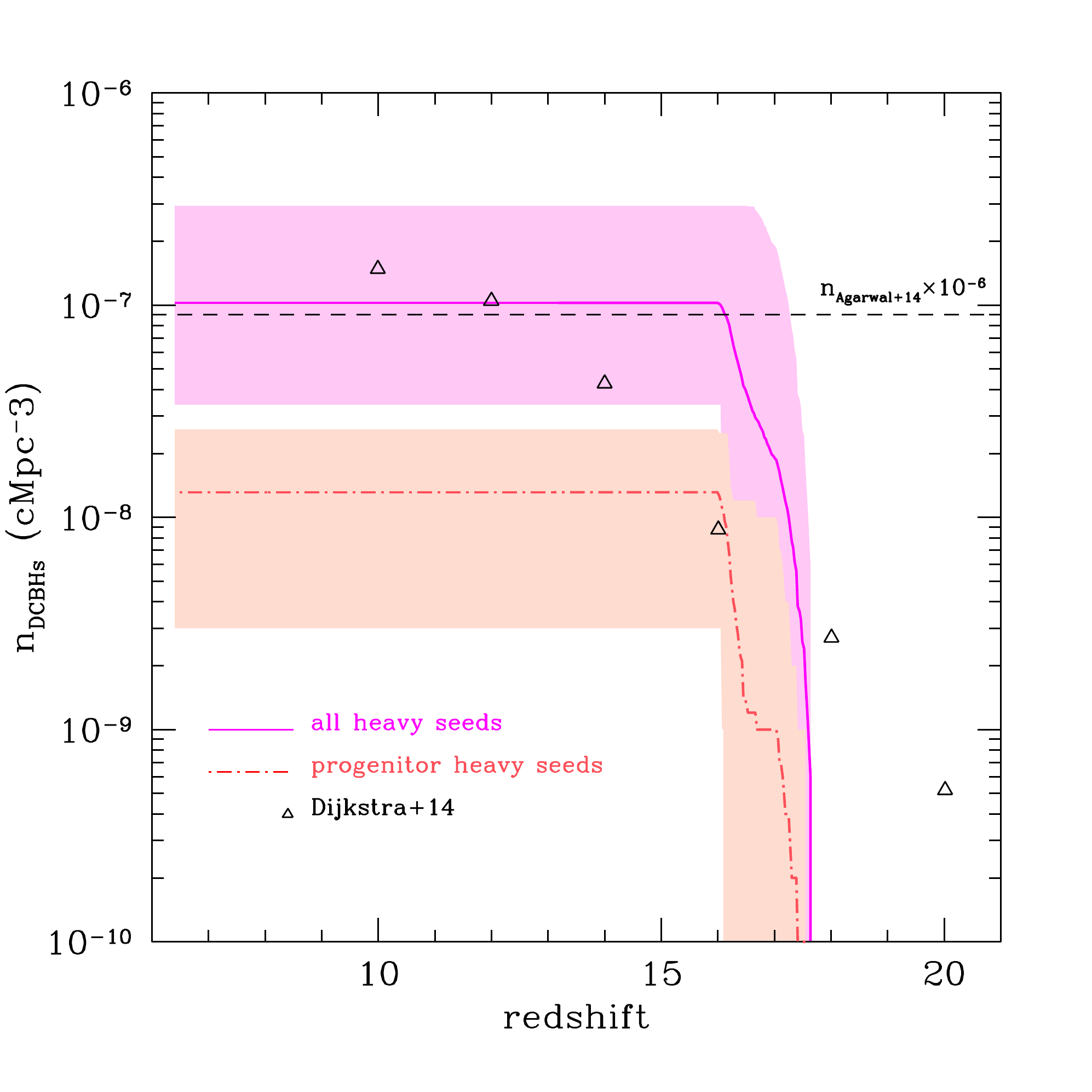}
\caption{Predicted number density of DCBHs as a function of $z$.  
The solid line shows the average number of all DCBHs formed in the reference model and the dot-dashed line 
shows the same quantity for real DCBH progenitors (see text). For each of these two classes, the shades span 
the values found in different merger tree simulations. For comparison, we also show the values predicted by
Dijkstra et al. (2014, traingles) and Agarwal et al. (2014, horizontal dashed line). The latter value has been 
multiplied by $10^{-6}$ to enable the comparison.} 
\label{fig:DCBHdens} 
\end{figure} 

We conclude that  the formation of a $> 10^9 \, \rm M_\odot$ BH at $z > 6$ depends on a
complex interplay of feedback processes, where the mass and redshift distribution
of both light and heavy seeds have a fundamental role.  
The first SMBHs can grow by Eddington-limited
accretion only if sufficiently massive BH seeds are able to form in their progenitor galaxies. This can
be achieved by means of  Pop~III BH remnants (light seeds) if  Pop~III stars form with $m_\ast > 300 \, \rm M_\odot$
or if smaller mass BH remnants merge to form a single, more massive BH.
Alternatively, even a few heavy seeds with mass $\sim 10^5 \, \rm M_\odot$ can provide the right ``head-start'', but their formation requires favourable conditions that can only be achieved if $J_{\rm cr} \lesssim 300$
and $Z_{\rm cr} \geq 10^{-3.8}$ Z$_\odot$. Since $J_{\rm cr}<300$ is lower than required by 3D cosmological simulations of the collapse of primordial clouds, alternative models of direct collapse driven by dynamical processes should be kept in mind (e.g. \citealt{LoebRasio94, EisensteinLoeb95, Begelman06, Mayer07, VB10}).

In our reference model, where the formation of a SMBH relies on heavy seeds, the  number of heavy seeds progenitors 
varies between 3 and 27 among the different merger tree simulations, with an average number of 13.
If we weight these numbers by the observed comoving density of quasars at $z = 6$, $n_{\rm SMBH} \sim 10^{-9} \,\, \rm{cMpc}^{-3}$, 
we can predict the comoving number density of direct collapse BHs (DCBHs). This is shown in 
Fig.~\ref{fig:DCBHdens}, where we compare the results of our reference model with other studies. We find that by $z \sim 15$ the
comoving number density of DCBH is $3 \times 10^{-9} \, \,  {\rm cMpc}^{-3} \lesssim n_{\rm DCBH} 
\lesssim 2.7 \times 10^{-8}  \, \, \rm{cMpc}^{-3}$. We clarify that these numbers refer to heavy BH seeds progenitors of SMBHs at $z = 6$.
In the same reference model, a much larger population of direct collapse BHs form ($\sim 100$ on average) which - however - end up as 
satellites and do not directly contribute to the mass growth of the SMBH. For these enlarged population, the comoving number density
that we predict is $\sim 10^{-7} \, \, \rm{cMpc}^{-3}$, in very good agreement with the results of \citet{Dijkstra14} for their fiducial model.
A much larger value has been found by \citet{Agarwal14}, who report $\sim 6$ potential DCBH hosts in their 4 cMpc size simulation box 
at $z \sim 10$, leading to an estimated comoving number density of $\sim 0.09  \, \, \rm{cMpc}^{-3}$. As already noted by \citet{Dijkstra14},
the main reason for this discrepancy is that \citet{Agarwal14} adopts a $J_{\rm cr} = 30$. No DCBH would form in their box if 
$J_{\rm cr} = 300$ were to be assumed (see the bottom panel of their Fig.~1). 

Compared to these previous analyses, our study allows to identify BH seeds that are the progenitors of the first
SMBHs, and to study the conditions that allow these BH seeds to germinate.

\section*{Acknowledgments}
We thank the anonymous referee for useful comments and suggestions.
The research leading to these results has received funding from the European Research Council under the European Union's
Seventh Framework Programme (FP/2007-2013) / ERC Grant Agreement n. 306476. 
This work is supported in part by the Grant-in-Aid from the Ministry of Education, Culture,
Sports, Science and Technology (MEXT) of Japan (25287040 KO).

\appendix

\section{Halo gas cooling efficiency}

In each dark matter halo, the fraction of gas mass that is able to cool is set by the balance 
between the cooling time and the dynamical time. Here we adopt a procedure similar to
that presented in \citet{M01}. 

We compute the free-fall time as,
\begin{equation}
t_{\rm ff}(r) = \int_0^r \frac{dr'}{\sqrt{v^2_{\rm e}(r')-v^2_{\rm e}(r)}} \left( = \int_0^r \frac{dr'}{v_r(r')}\right),
\end{equation}
\noindent
where $v_{\rm e}$ is the escape velocity ($v_r$ is the infall velocity of a test particle at rest at $r$), 
dark matter halos are assumed to have a NFW density profile with concentration parameter $c = 4.8$,
and the gas follows an isothermal gas density distribution
(see eqs. 9, 18 and 21 in \citealt{M01}). 

The cooling time is defined as,
\begin{equation}
t_{\rm cool} (r) = \frac{3 \, n \, k\, T_{\rm vir}}{2\, \Lambda(n, Z)},
\end{equation}
\noindent
where $k$ is Boltzmann's constant, $n$ is the gas number density and $\Lambda(n, Z)$ is the density and
metallicity dependent cooling rate per unit volume. 

The parameter $f_{\rm cool}$ introduced in section \ref{sec:SF} is defined as the ratio between the gas mass
within a radius $r_{\rm cool}$ such that $t_{\rm cool}(r_{\rm cool}) = t_{\rm ff}(r_{\rm cool})$ and the total gas
mass within the virial radius. 

In Lyman-$\alpha$ halos, even in primordial conditions the gas can efficiently cool by means of H and He transitions.
Hence, we assume that in these systems $f_{\rm cool} = 1$ and the SFR is only limited by the infall rate (see eq.~\ref{eq:sfr})\footnote{This condition is strictly true only at $z \ge 9$ for halos with $T_{\rm vir} < 5 \times 10^5$~K.
In fact, above this temperature and in primordial conditions the cooling rate is dominated by free-free emission. Here
we assume $f_{\rm cool} = 1$ for all Lyman-$\alpha$ halos because most of these larger virial temperatures (mass) halos
are already metal-enriched when they first appear along the merger tree and therefore the cooling rate is dominated by
highly ionized metal species \citep{SD93}.}.

In mini-halos the cooling time
 can be longer than the free-fall time at most radii, so that $f_{\rm cool} \ll 1$. The 
exact value of this parameter depends on the virial temperature, redshift (hence halo mass) and metallicity of 
the gas. The cooling rate is computed considering a simplified version of the chemical evolution model of
\citet{Omukai2012} that we describe below.  \\

\noindent
{\it Cooling and heating processes:}\\
To compute the cooling rate we consider the following physical processes: H Ly$\alpha$
emission ($\Lambda_{\rm Ly\alpha}$), H$_2$ rovibrational emission ($\Lambda_{\rm H_2}$), and CII and OI
fine-structure line emission ($\Lambda_{\rm CII}$, $\Lambda_{\rm OI}$). Photoelectric emission
by dust ($\Gamma_{\rm PE}$) is also taken onto account as it provides an important heating process when
the medium is dust enriched and in the presence of a far UV (FUV) background (Wolfire et al. 1995). For
simplicity, we assume the same spectral shape in the Galactic ISM and we take the Habing parameter for
the FUV to be $G_0 = 2.9 \times 10^{-2} J_{\rm LW}$. Hence, the total cooling rate is computed as:
\begin{equation}
\Lambda = \Lambda_{\rm Ly\alpha} + \Lambda_{\rm H_2} + \Lambda_{\rm CII} + \Lambda_{\rm OI} - \Gamma_{\rm PE}.
\end{equation}

\noindent
{\it Ionization degree:}\\
The post-recombination leftover electron fraction is $y_{\rm e, prim} \sim 2 \times 10^{-4}$. Following
virialization, the ionization degree can increase due to collisional ionization: 
\begin{equation}
\rm H + e \rightarrow H^{+} + 2 e
\end{equation}
\noindent
followed by radiative recombination:
\begin{equation}
\rm H^+ + e \rightarrow H + \gamma.
\end{equation}
\noindent
Hence, the ionization fraction reaches an equilibrium value given by:
\begin{equation}
y_{\rm e, eq} = \frac{k_{\rm ion}}{k_{\rm ion}+k_{\rm rec}}.
\end{equation}
\noindent
In the model, we take the ionization degree to be $y_{\rm e} = {\rm max}(y_{\rm e, prim}; y_{\rm e, eq})$ and the 
atomic hydrogen fraction as $y_{\rm H} \sim 1 -y_{\rm e}$ because the molecular fraction is always $\ll1$.\\

\noindent
{\it Molecular fraction:}\\
Molecular hydrogen can form from the gas phase via the H$^-$ channel:
\begin{eqnarray}
\rm H + e \rightarrow H^- + \gamma \\
\rm H^- +H \rightarrow H_2 + e
\end{eqnarray}
\noindent
or on the surface of dust grains. In one free-fall time, the H$_2$ fraction formed can be approximated as:
\begin{equation}
y_{\rm H_2, form} =\frac{k_{\rm H^-, form}}{k_{\rm rec}} \, {\rm ln} \left (1+\frac{t_{\rm ff}}{t_{\rm rec}}\right) + k_{\rm dust, form} \, y_{\rm H} \, n \, t_{\rm ff},
\end{equation}
\noindent
where $k_{\rm H^-, form} \, n_{\rm H} \, n_{\rm e}$ is the $\rm H_2$ formation rate via  the H$^-$ channel, $k_{\rm dust, form}$ is the $\rm H_2$
formation rate on dust grains, and the recombination time is:
\begin{equation}
t_{\rm rec} = \frac{1}{ k_{\rm rec} \, y_{\rm e} \, n}.
\end{equation}
\noindent
In the presence of a LW background, the $\rm H_2$ fraction reaches an equlibrium value given by \citep{ag89}:
\begin{equation}
y_{\rm H_2, eq}  = \left(k_{\rm H^-, form} \, y_{\rm e} + k_{\rm dust, form}\right)\, \frac{y_{\rm H} \, n}{k_{\rm dis}},
\end{equation}
\noindent
where $k_{\rm dis}$ is the $\rm H_2$ dissociation coefficient and it is calculated considering the $\rm H_2$ self-shielding factor as is 
\citet{WH11}. In the model, we take the $\rm H_2$ fraction to be $y_{\rm H_2} = {\rm min} (y_{\rm H_2, form}; y_{\rm H_2, eq})$. \\

\noindent
{\it Metal fractions:}\\
For simplicity, at each given metallicity $Z$ all the carbon atoms are assumed to be in CII and the oxygen atoms in OI,
with an elemental abundance given by:
\begin{equation}
y_{\rm CII} = 3.97 \times 10^{-4} \, \frac{Z}{\rm Z_\odot} \,\,\,\,\,\,\,\,\,\, y_{\rm OI} = 8.49 \times 10^{-4}  \, \frac{Z}{\rm Z_\odot}.
\end{equation}
\noindent
When dust grains are present, we account for partial depletion of these two elements on dust grains and we assume
an elemental fraction of \citep{Pollack94}:
\begin{equation}
y_{\rm CII} = 0.93 \times 10^{-4} \, \frac{Z}{\rm Z_\odot} \,\,\,\,\,\,\,\,\,\, y_{\rm OI} = 3.57 \times 10^{-4}  \, \frac{Z}{\rm Z_\odot}.
\end{equation}
\noindent
\\

\begin{figure}
\includegraphics [width=8.0cm]{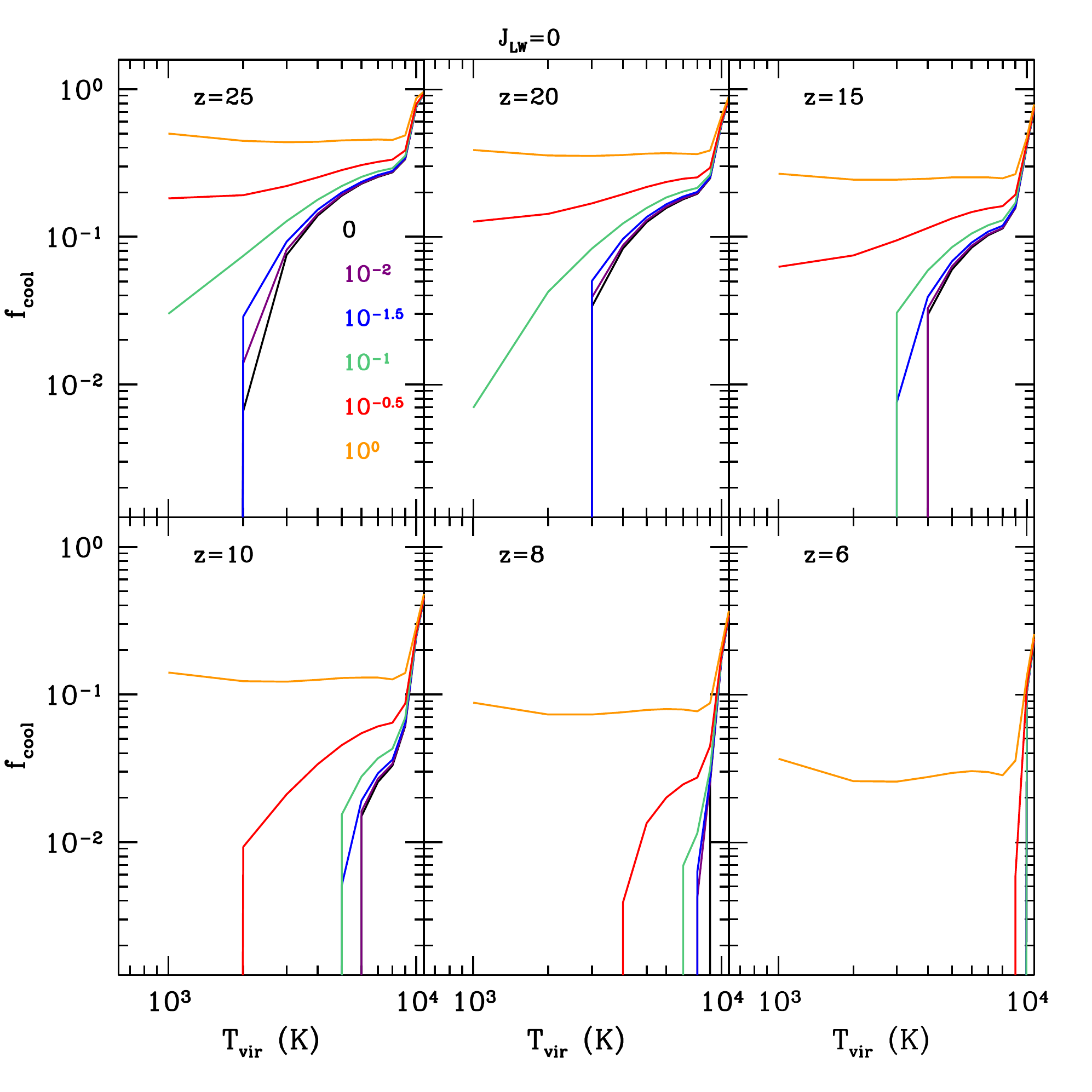}
\caption{The mass fraction of gas that is able to cool in one free-fall time, $f_{\rm cool}$ as a function of halo virial
temperature for $J_{\rm LW}=0$. Each line represents a fixed gas metallicity: $Z = 0$ (black), $10^{-2}$ (violet), $10^{-1.5}$ (blue), $10^{-1}$ (green), 
$10^{-0.5}$ (red), $1$ (yellow), in solar units. Here we consider the presence of dust grains (see text). Different panels are for 
different redshift, from $z=25$ to $z=6$, as labelled in the figure.}
\label{fig:A1} 
\end{figure}

\begin{figure}
\includegraphics [width=8.0cm]{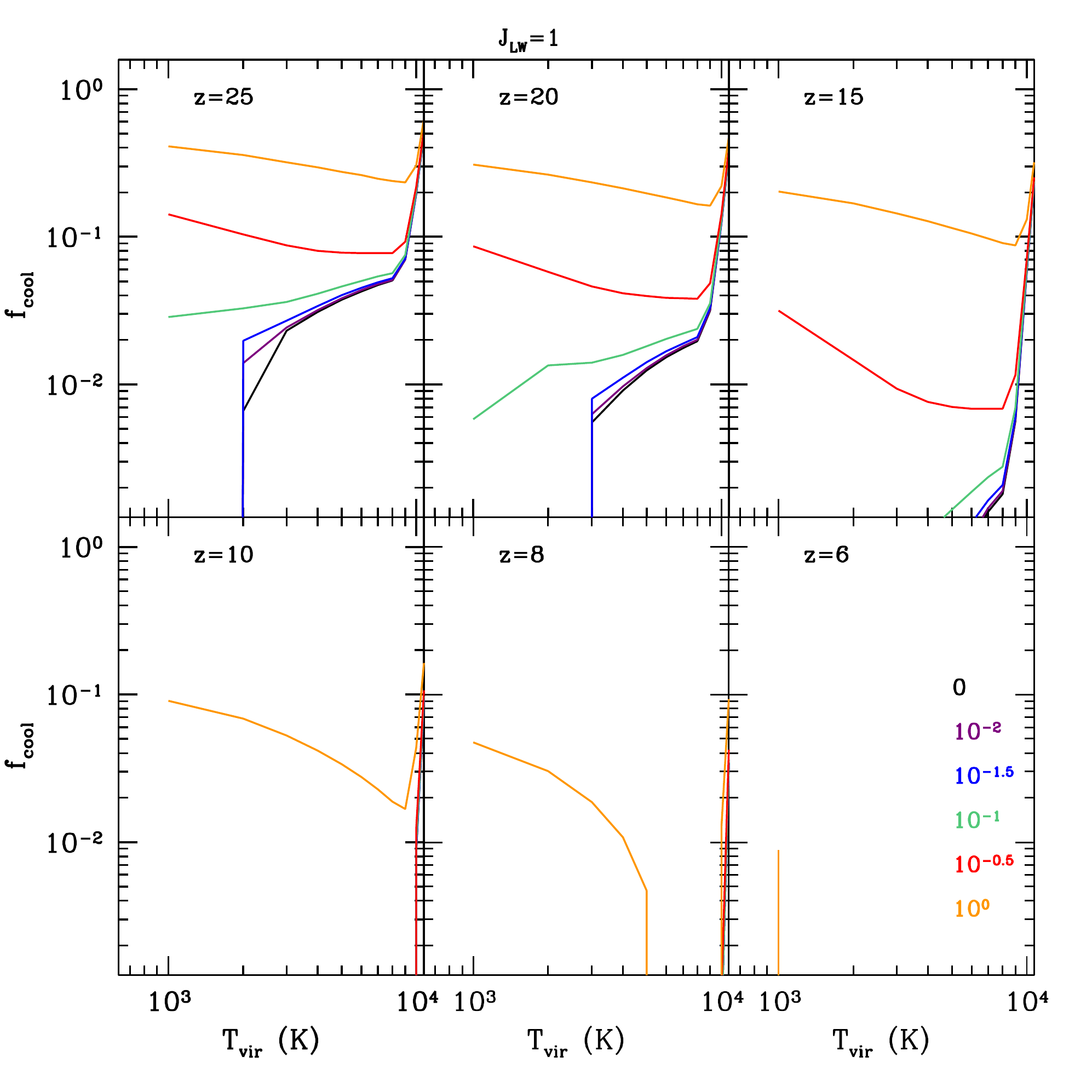}
\caption{Same as Fig.~\ref{fig:A1} but for $J_{\rm LW}=1$.}
\label{fig:A2} 
\end{figure}

\begin{figure}
\includegraphics [width=8.0cm]{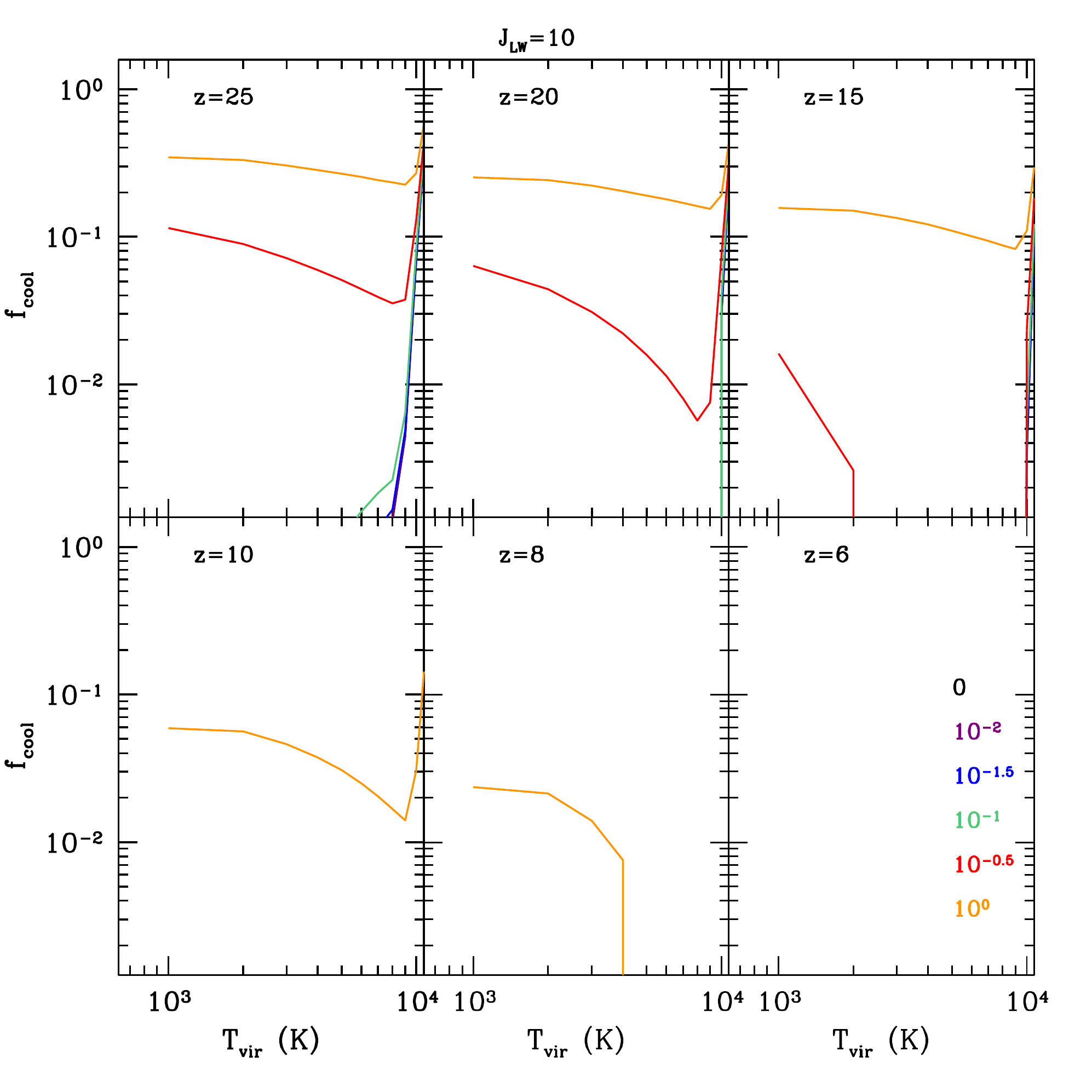}
\caption{Same as Fig.~\ref{fig:A1} but for $J_{\rm LW}=10$.}
\label{fig:A3} 
\end{figure}

\begin{figure}
\includegraphics [width=8.0cm]{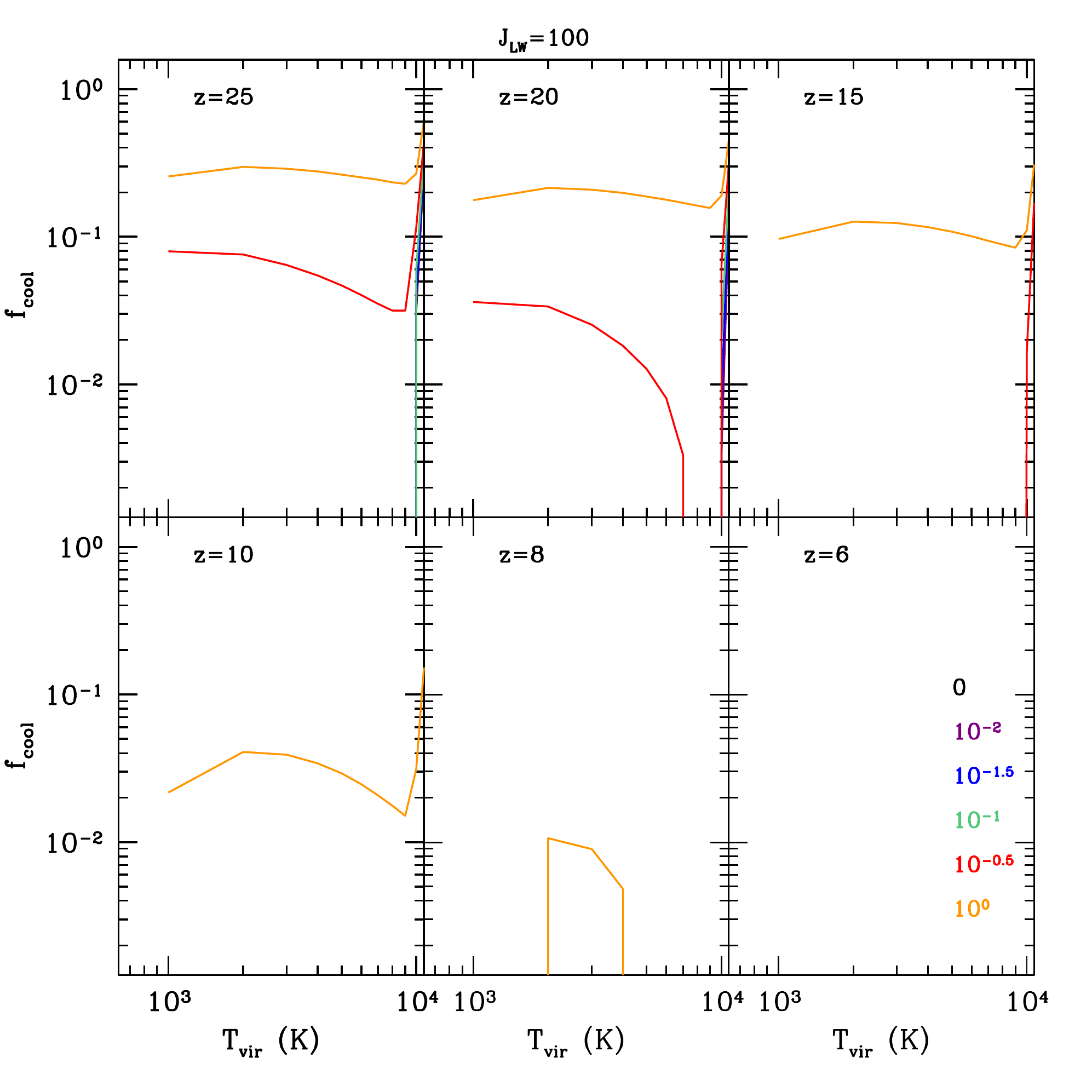}
\caption{Same as Fig.~\ref{fig:A1} but for $J_{\rm LW}=100$.}
\label{fig:A4} 
\end{figure}

Using the above prescription we run a large set of models, changing the virial temperature, redshift, gas metallicity, and the value of the external
LW background, and we compute the corresponding $f_{\rm cool}$. The results are summarized in Figs.~\ref{fig:A1}-\ref{fig:A4}. When $Z = 0$,
$f_{\rm cool}$ is a strong function of $T_{\rm vir}$. In the absence of a LW background (see Fig.\ref{fig:A1}), $f_{\rm cool}$ drops from $\sim 1$
to $\sim 0.1$ in the temperature range $8000~{\rm K} \le T_{\rm vir} < 10^4~{\rm K}$, followed by a smoother decline to $\sim 10^{-2}$ at 
$T_{\rm vir} \sim 10^3~{\rm K}$. Even in these favourable conditions (no $\rm H_2$ dissociation), for a fixed $T_{\rm vir}$ 
$f_{\rm cool}$ strongly depends on redshift. This reflects the density dependence of the cooling rate, which - for the same physical
conditions - leads to a shorter cooling time at high redshift. This result holds even when the gas metallicity is $Z > 0$. Significant deviations
from the primordial case require a metallicity of $Z \ge 0.1$ Z$_{\odot}$ at $z \gtrsim 20$, of $Z \ge 0.5$ Z$_\odot$ at $10 \lesssim z \lesssim 15$,
of $Z \ge$ Z$_\odot$ at $6 \lesssim z \lesssim 8$. Finally at $z < 6$ we find that - independently of the gas metallicity - gas cooling in mini-halos is suppressed.

Figs.\ref{fig:A2}-\ref{fig:A4} show how the above results change when the intensity of the LW background ranges between $1 \le J_{\rm LW} \le 10^2$. 
As expected, the largest effect is in the behaviour of the $Z < 0.1$ Z$_\odot$ gas, which can cool only if $J_{\rm LW} \lesssim 1$ and $z \gtrsim 20$.

\bibliography{biblioTest}
\label{lastpage}

\end{document}